\documentclass[superscriptaddress,aps,prl,twocolumn]{revtex4-1}
\usepackage{amssymb}
\usepackage{amsmath}
\usepackage[dvipsnames]{xcolor}
\usepackage{chngpage}
\usepackage{lgrind}        
\usepackage{color}         
\usepackage{graphics}      
\usepackage[pdftex]{graphicx}      
\usepackage{longtable}     
\usepackage{epsf}          
\usepackage{bm}            
\usepackage{asymptote}     
\usepackage{thumbpdf}
\usepackage{verbatim}
\usepackage{url}
\usepackage{MnSymbol}
\usepackage{physics}
\usepackage[colorlinks=true]{hyperref} 
\usepackage{enumitem}	

\usepackage{float}
\usepackage{times}
\usepackage{color}
\usepackage{verbatim}
\usepackage{soul,xcolor}
\setstcolor{red}
\usepackage{colortbl}

\usepackage{tabularx}

\usepackage[normalem]{ulem} 

\newcommand{\outline}[1]{}   

\makeatletter
\AtBeginDocument{
\g@addto@macro{\appendix}{\renewcommand{\p@subsection}{\@Alph\c@section}}
}
\makeatother

\begin{document}
\title{
Controlling Local Thermalization Dynamics in a Floquet-Engineered Dipolar Ensemble}

\author{Leigh S. Martin$^{1,\ast}$, Hengyun Zhou$^{1,\ast,\dagger}$, Nathaniel T. Leitao$^{1,\ast}$, Nishad Maskara$^1$, Oksana Makarova$^{1,2}$, Haoyang Gao$^1$, \\Qian-Ze Zhu$^{1,2}$, Mincheol Park$^1$, Matthew Tyler$^1$, Hongkun Park$^{1,3}$, Soonwon Choi$^4$, Mikhail D. Lukin$^{1,\dagger}$\\
\normalsize{$^1$Department of Physics, Harvard University, Cambridge, Massachusetts 02138, USA}\\
\normalsize{$^2$School of Engineering and Applied Sciences, Harvard University, Cambridge, Massachusetts 02138, USA}\\
\normalsize{$^3$Department of Chemistry and Chemical Biology, Harvard University, Cambridge, Massachusetts 02138, USA}\\
\normalsize{$^4$Center for Theoretical Physics, Massachusetts Institute of Technology, Cambridge, Massachusetts 02139, USA}\\
\normalsize{$^\ast$These authors contributed equally to this work.}\\
\normalsize{$^\dagger$To whom correspondence should be addressed; E-mail: hzhou@g.harvard.edu,  lukin@physics.harvard.edu}}

\begin{abstract}
    Understanding the microscopic mechanisms of thermalization in closed quantum systems is among the key challenges in modern quantum many-body physics.
    We demonstrate a method to probe local thermalization in a large-scale many-body system by exploiting its inherent disorder, and use this to  uncover the thermalization mechanisms in a three-dimensional, dipolar-interacting spin system with tunable interactions.
    Utilizing advanced Hamiltonian engineering techniques to explore a range of spin Hamiltonians,
    we observe a striking change in the characteristic shape and timescale of local correlation decay as we vary the engineered exchange anisotropy. We show that these observations originate from the system's intrinsic many-body dynamics and reveal the signatures of conservation laws within localized clusters of spins, which do not readily manifest using global probes.
    Our method provides an exquisite lens into the tunable nature of local thermalization dynamics, and enables detailed studies of scrambling, thermalization and hydrodynamics in strongly-interacting quantum systems.
\end{abstract}

\maketitle

\outline{Introduction}
Thermalization in isolated quantum many-body systems underlies the emergence of quantum statistical mechanics.
This happens despite the unitary, reversible evolution of a closed quantum system, and is commonly understood from the perspective of the system acting as its own bath, as formalized by the eigenstate thermalization hypothesis~\cite{dalessio2016from,abanin2019colloquium,langen2015experimental,schreiber2015observation,kaufman2016quantum,kucsko2018critical,wei2018exploring,choi2019probing}.
Equally important is the dynamics by which a system reaches thermal equilibrium.
Recent work has uncovered various universal phenomena, including scrambling~\cite{hayden2007black,shenker2014black,landsman2019verified} and hydrodynamic transport~\cite{wei2022quantum,jepsen2020spin,zu2021emergent,bulchandani2021superdiffusion}.
However, many aspects of this approach to thermal equilibrium are still poorly understood, particularly in regards to how thermalizing dynamics and the eigenstate thermalization hypothesis emerge from coherent interactions within a closed system.

In this Letter, we demonstrate a new tool to probe local dynamics in strongly-interacting systems without the need for single-spin control or readout. We apply it to the paradigmatic XXZ model in a positionally disordered, dipolar spin system, in which the nanometer-scale spin-spin separation and three-dimensional geometry make single-site operations infeasible.
Combining this method with advanced Hamiltonian engineering pulse sequences~\cite{choi2020robust,zhou2023robust,tyler2023higher,wei2018exploring,geier2021floquet}, we transform the native system Hamiltonian into a wide range of XXZ Hamiltonians and access qualitatively distinct regimes of equilibration.
We find that the local thermalization dynamics are consistent with coherently coupled clusters of spins that interact with
each other via fluctuating magnetic fields, whose correlation times and hybridization determine the timescale and shape of the decay.
Our method provides a powerful lens into the tunable nature of local relaxation dynamics in closed quantum many-body systems, which is not accessible via global Ramsey probes~\cite{geier2021floquet}.

\begin{figure}
\begin{center}
\includegraphics[width=\columnwidth]{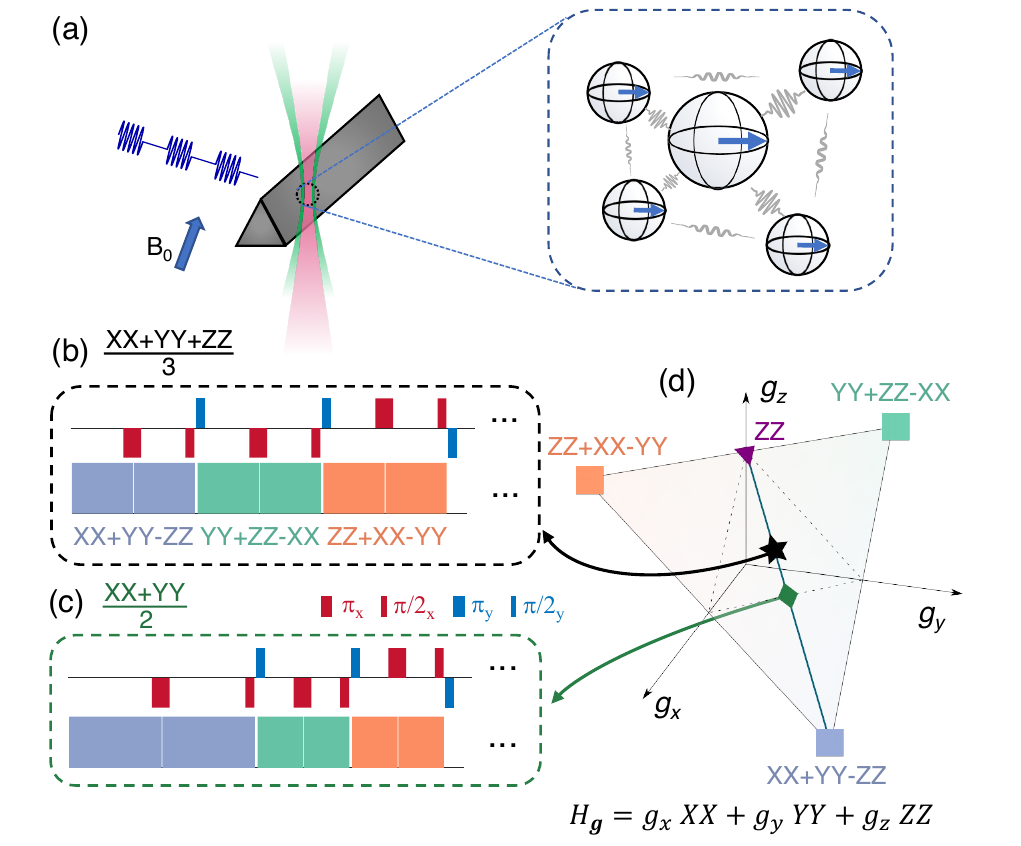}
\caption{{\bf Experimental system and Hamiltonian engineering.} (a) Black diamond experimental system, consisting of a high density ensemble of NV spins in diamond. Spin initialization and readout are achieved via optical illumination and fluorescence, while spin manipulation is performed via microwave pulses. (b,c) Representative pulse sequence block and illustration of Hamiltonian engineering concept. By tuning pulse separations in an interaction decoupling sequence, the effective Hamiltonian can be engineered into different forms, such as Heisenberg (b) or XY (c) Hamiltonians. (d) Illustration of accessible XYZ Hamiltonians via Hamiltonian engineering. The accessible Hamiltonians are averages of those obtained via transforming the native Hamiltonian by a $\pi/2$ pulse [squares]. Special Hamiltonians of interest are labeled: Ising [purple triangle], Heisenberg [black star], XY Hamiltonian [green diamond].} 
\label{fig:fig1}
\end{center}
\end{figure}

\outline{Experimental System and Dynamical Hamiltonian Engineering}
\emph{Experimental system and Hamiltonian engineering.}---
Our experimental system (Fig.~\ref{fig:fig1}(a)) consists of a high density ($\sim$15 ppm), positionally-disordered ensemble of negatively-charged nitrogen-vacancy (NV) centers in diamond~\cite{doherty2013nitrogen}.
The NV center spin in its electronic ground state forms a spin-1 triplet, from which we isolate a spin-1/2 degree of freedom via the application of an external magnetic field aligned with one group of NVs with the same lattice orientation.
The high density of NV centers enables strong, dipolar interactions between nearby spins ($J\approx (2\pi)35$ kHz at a typical separation).
Additional paramagnetic defects and lattice strain result in large on-site disorder ($W\approx (2\pi)4 $ MHz) of the spins.
Green laser illumination enables optical polarization of the spin state, while fluorescence on the red phonon sideband allows read out of the final global spin polarization.
Microwave pulses resonant with the target NV spin transition frequency allow fast manipulation of the spin state and Hamiltonian engineering (see Ref.~\cite{SM,kucsko2018critical,zhou2020quantum} for experimental details).

We utilize an improved version~\cite{zhou2023robust,tyler2023higher} of the robust Hamiltonian engineering techniques (Fig.~\ref{fig:fig1}(b,c)) introduced in Ref.~\cite{choi2020robust} to suppress local disorder and engineer tunable XYZ spin-spin interaction Hamiltonians
\begin{align}
    H_{\bm g} = \sum_{ij} J_{ij} \left(g_x S_i^x S_j^x+g_y S_i^y S_j^y+g_z S_i^z S_j^z\right),
\end{align}
parameterized by an anisotropy vector, $\bm g=(g_x, g_y, g_z)$.
This is accomplished by tuning the evolution time along each of the three axis directions (Fig.~\ref{fig:fig1}(b,c)). 
Here, $J_{ij} \propto 1/r^3_{ij}$ is the long-ranged, anisotropic dipolar interaction strength.
We  primarily focus on XXZ Hamiltonians, where we parameterize the interaction as 
$\bm g(\lambda_\textrm{XXZ})= \left(1+\lambda_\textrm{XXZ}, 1+\lambda_\textrm{XXZ}, 1-2\lambda_\textrm{XXZ} \right)/3$ with $\lambda_\textrm{XXZ}$ characterizing the distance away from the $SU(2)$ symmetric Heisenberg Hamiltonian, see Fig. \ref{fig:fig1}(d).
Although our main pulse sequence cannot engineer the Ising point $\bm{g} = (0,0,1)$ due to finite-pulse effects, we use a spin locking sequence (i.e. continuous driving) to access it.

\outline{Measuring Global and Local Thermalization}
\emph{Global and local probes of thermalization.}---
Equipped with the ability to engineer various XXZ Hamiltonians, we now explore the thermalization dynamics.
We first examine the dynamics of global observables, utilizing a conventional Ramsey sequence to measure the decay of a polarized initial state along $+\hat{x}$~\cite{nandkishore2021lifetimes}, given by
\begin{align}
    \mathcal{C}^{XX}_{\text{Global}}(t) = \frac{2}{N}\langle S^x(t)\rangle_{+\hat{x}}= \left(\frac{2}{N}\right)^2 \langle S^x(t)S^x(0)\rangle_{+\hat{x}},\label{eq:ramsey}
\end{align}
where $S^\mu(t) = \sum_i S_i^\mu(t) = \sum_i e^{i H_{\bm{g}}t} S_i^\mu e^{-i H_{\bm{g}}t}$ is the Heisenberg picture \textit{global} spin operator and $\langle\rangle_{+\hat{x}}$ denotes an ensemble average in the state where all spins are initially polarized along the $+\hat{x}$ direction.

\begin{figure}
\begin{center}
\includegraphics[width=\columnwidth]{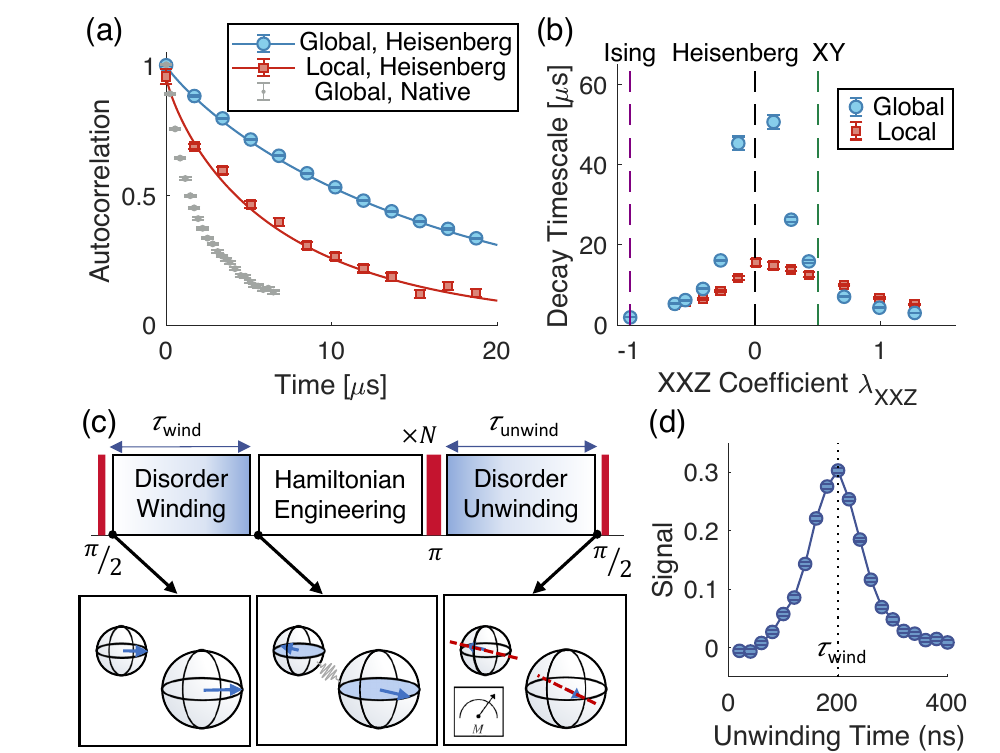}
\caption{{\bf Measuring global and local spin autocorrelators.}
(a) Measured coherence decay under an engineered Heisenberg Hamiltonian for global [blue circles] and local [red squares] spin autocorrelators. Fit is a stretched exponential. For reference, we also show the XY-8 decay [gray points], which characterizes the decay under the native interaction Hamiltonian.
(b) Decay times for normalized global and local spin autocorrelators. Vertical dashed lines denote special Hamiltonians of interest.
(c) ``Disorder Order" sequence to measure the decay of local spin autocorrelators for an infinite temperature initial state, which reveals local thermalization.
The sequence consists of a $\pi/2$ initialization pulse, a free evolution duration to encode the local disorder strength into the spin state, a varying number of repetitions of the Hamiltonian engineering sequence, followed by a $\pi$ pulse and free evolution to rephase the spins and $-\pi/2$ pulse for spin state readout.
The sequence is designed to suppress higher-order corrections (see Ref.~\cite{SM} for details).
(d) Measured normalized spin polarization as a function of disorder unwinding time, showing a revival when the winding and unwinding times are equal.
In this example, the Hamiltonian engineering consists of two Floquet cycles of Heisenberg Hamiltonian engineering.
}
\label{fig:fig2}
\end{center}
\end{figure}

As a validation of our Hamiltonian engineering tools, we first engineer the Heisenberg Hamiltonian $\lambda_{XXZ}=0$, which exhibits a global SU(2) symmetry and therefore preserves uniformly polarized initial states. As seen in Fig.~\ref{fig:fig2}(a), global Heisenberg dynamics display an order of magnitude longer decay timescale than the disorder-decoupled native NV interaction Hamiltonian (gray points).
To characterize the timescale and shape of the decay, we fit the signal to a stretched exponential form $\mathcal{C}(t)\propto \exp(-(t/\tau)^\nu)$, where $\tau$ describes the characteristic timescale, and the stretching exponent $\nu$ encodes the shape (solid lines in Fig.~\ref{fig:fig2}(a)).
In Fig.~\ref{fig:fig2}(b), the blue points show the decay timescale for a range of different XXZ Hamiltonians, normalized by the decay curve at the Heisenberg Hamiltonian, which is completely dominated by extrinsic factors.
Conversely, the prominent peak around the Heisenberg Hamiltonian confirms that the normalized decay is dominated by dynamics of the engineered Hamiltonian.

While the Heisenberg Hamiltonian freezes the decay of any polarized initial state, it still induces dynamics in generic initial states, leading to local thermalization.
To probe this local equilibration for generic initial states, we introduce a technique to measure the infinite temperature \textit{local} spin autocorrelators, despite only having access to native \textit{global} control and measurements, by leveraging the inherent large disorder of the system.
We refer to this probe as a ``disorder-order" measurement~\cite{jeener1967nuclear}. It prepares the spins in a random product state encoding the local disorder strength on each spin to mimic an infinite temperature quench.

The measurement sequence is illustrated in Fig.~\ref{fig:fig2}(c), and resembles the familiar spin-echo technique.
The disorder-winding and unwinding free evolutions surrounding the Floquet Hamiltonian engineering serve two essential purposes.
First, the initial free-evolution step distributes each spin uniformly along the equator of its Bloch sphere, with the disorder field $h_i$ imprinting a local phase $\theta_i = h_i \tau_{\text{wind}}$. This step initializes a random product state at infinite effective temperature.
Second, reversing the initial disorder-winding prior to measurement of global polarization transforms the spatially homogenous measurement of the ensemble into a spatially inhomogenous measurement, where each spin is locally measured along the direction in which they were initially prepared.
This local realignment (see Fig.~\ref{fig:fig2}(d)) allows only the local operator autocorrelations associated to the plane of the disorder-winding to survive the disorder average. The resulting signal is
\begin{align}
    \frac{2}{N}\sum_i \langle S_i^x(t) S_i^x(0)\rangle_{T=\infty}+\langle S_i^y(t) S_i^y(0)\rangle_{T=\infty},
    \label{eq:do}
\end{align}
where $\langle \rangle_{T=\infty}$ denotes an expectation value over the infinite temperature state~\cite{SM}.
Note that despite their superficial similarities, Eq.~(\ref{eq:do}) is very different from Eq.~(\ref{eq:ramsey}); for example, a Heisenberg Hamiltonian does not cause any decay in Eq.~(\ref{eq:ramsey}), but does cause local equilibration in Eq.~(\ref{eq:do}).
We further generalize this protocol for disordered rotations around the $\hat{x}$, $\hat{y}$, $\hat{z}$ axes and combine the results to infer the local autocorrelations of each axis individually, \textit{i.e.} $ \mathcal{C}^{\mu \mu}_{\text{Local}}(t)  = \frac{4}{N}\sum_i \langle S_i^\mu(t) S_i^\mu(0)\rangle_{T=\infty}$ for $\mu = X,Y,Z.$

The decay timescales for the disorder-order measurements are shown as the red traces in Fig.~\ref{fig:fig2}.
In Fig.~\ref{fig:fig2}(a,b), we find that the decay of such local correlators for the Heisenberg Hamiltonian is significantly faster than the global correlators, confirming that the disorder-order technique detects local thermalization, even in the SU(2)-symmetric case that preserves polarized initial states.

\begin{figure}
\begin{center}
\includegraphics[width=\columnwidth]{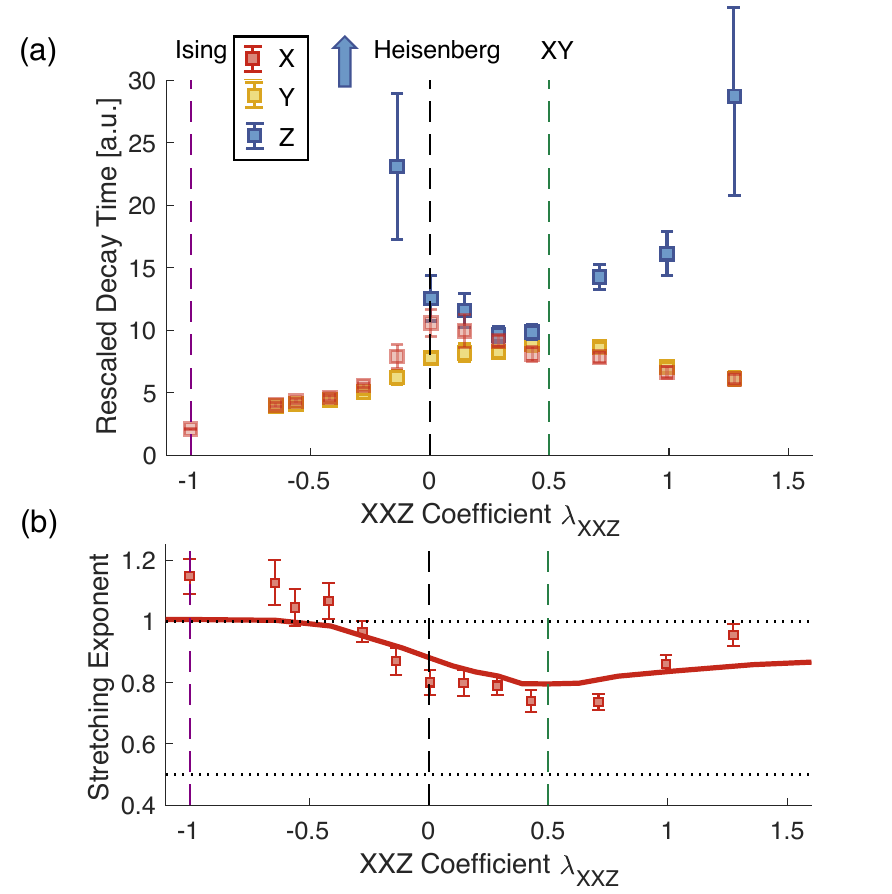}
\caption{{\bf Decay timescale and decay shape for local correlators.} (a) Local X [red], Y [yellow] and Z [blue] spin autocorrelator decay timescales, normalized by extrinsic decay and rescaled by Hamiltonian norm.
The blue arrow indicates divergent timescales of Z correlators closer to the Ising Hamiltonian~\cite{SM}.
(b) Local spin autocorrelator decay shapes, averaged over X, Y, as characterized by the stretching exponent.
Horizontal dotted line at $d/\alpha=1$ is the naive expectation based on existing arguments in the NMR literature, predicting an exponential decay.
A second dashed line at $0.5=d/2\alpha$ is also plotted as the expectation for dynamics which generate Markovian fluctuating fields.
Solid line is the prediction from a dynamical mean-field model.}
\label{fig:fig3}
\end{center}
\end{figure}

\outline{Diagnosing Thermalization Mechanisms}
\emph{Tuning local thermalization.}---
Interestingly, different XXZ Hamiltonians show markedly different decay shapes, as shown in Fig.~\ref{fig:fig3}(b).
Varying the XXZ Hamiltonian for local spin autocorrelators along X, we find a significant deviation of the decay shape from a simple exponential form, contrary to conventional NMR heuristics~\cite{Abragam1961}.
While the simple exponential shape qualitatively captures features close to the Ising Hamiltonian, we observe a striking decrease of the stretching exponent on the easy-plane side of the phase diagram $(\lambda_{XXZ}>0)$, where $|g_{x,y}|>|g_z|$. As we will see, the unexpected variations of the stretching exponent are an explicit manifestation of the intrinsic, quantum many-body noise being tuned by the effective Hamiltonian.

To provide a physical explanation for the observed decay timescales and shapes, we develop a simple physical model and complement it with numerical simulations that qualitatively reproduce the observations.

\begin{figure}
\begin{center}
\includegraphics[width=\columnwidth]{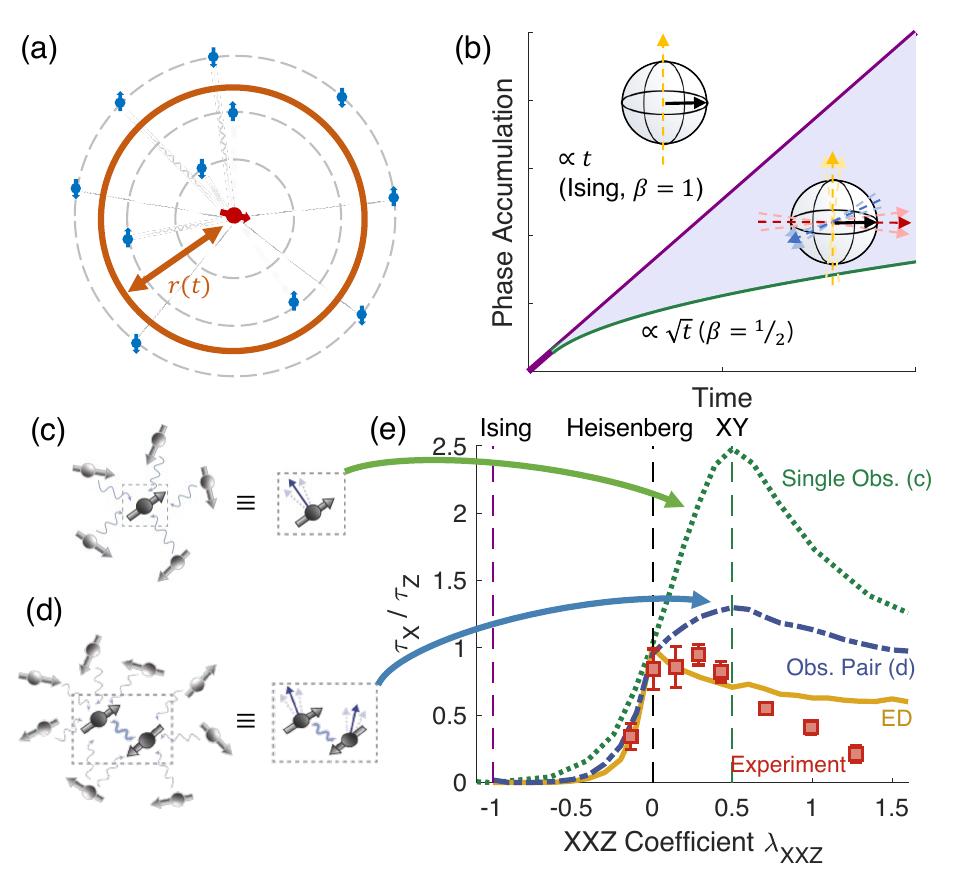}
\caption{{\bf Physical mechanism driving coherence decay.} (a) Quantum fluctuations of neighboring spins produce a dynamical mean-field that induces decay, and as time progresses, weakly coupled spins at larger distances start to contribute as well.
(b) For different Hamiltonians, the correlation time of the dynamical mean-field changes, resulting in different rates of phase accumulation over time (ballistic for a static field, diffusive for a Markovian fast-varying field) and leading to different decay shapes at early-to-intermediate times.
(c) Illustration of a single spin experiencing the fluctuating magnetic field from its neighbors.
(d) Illustration of the additional effect of hybridization between strongly coupled pairs of spins.
(e) Ratio between X and Z decay times of local correlators for experimental data and various models.
Including local hybridization [blue] improves the agreement with experimental data [red], and exact diagonalization of full quantum dynamics [yellow, 18 spins] agrees best.}
\label{fig:fig4}
\end{center}
\end{figure}

Focusing first on the decay shape of the engineered Ising dynamics, one can analytically calculate the global and local $S_x$ autocorrelators for randomly-positioned $d$-dimensional spin ensembles with power-law $1/r^\alpha$ Ising interactions~\cite{feldman1996configurational,davis2021probing,dwyer2021probing}.
For any given spin, quantum fluctuations of neighboring spins produce a random magnetic field $\hat{B}_i = \sum_j J_{ij} \hat{S}_j^z$ driving precession dynamics in the XY plane.
Crucially, this magnetic field is static, as the local magnetization $S_j^z$ is conserved under the ZZ Ising Hamiltonian, and spins precess ballistically as $\Delta\phi_t(r) \sim t/r^\alpha$. Counting the number of spins contributing to this precession, as depicted in Fig.~\ref{fig:fig4}(a), leads to a stretching exponent of $d/\alpha$, in agreement with our experimental observations.

Away from the Ising Hamiltonian, the non-zero flip-flop term $g_{x,y}$ transports magnetization through the spin bath, rendering the local field dynamical.
When the correlation time of the field becomes comparable to the precession timescale, the accrued phases can destructively interfere.
This qualitatively modifies the earlier scaling to $\Delta \phi_t(r) \sim t^\beta/r^\alpha$, where $1 \geq \beta \geq 1/2$ is a phenomenological parameter interpolating between ballistic and diffusive precession dynamics in the respective limits of static and Markovian fields, see Fig.~\ref{fig:fig4}(b). Shorter bath correlation times reduce the stretching exponent to $\nu = \beta d/\alpha$, and yield $d/2\alpha$ in the Markovian limit. 

To confirm that the bath correlation times determine the stretching exponents in the full non-commuting XXZ Hamiltonian, we use a minimal model that incorporates the correlation times and vector nature of the dynamic magnetic field into a self-consistent dynamical mean-field model~\cite{SM,graer2021dynamic}. In particular, we approximate the quantum magnetic fields by zero-mean, normally distributed classical variables whose temporal correlations are self-consistently determined by the local dephasing dynamics of neighbouring spins.
The resulting decay shapes and stretching exponents qualitatively match the experimental data (red line in Fig.~\ref{fig:fig3}(b)), validating our intuition that dynamic magnetic fields transverse to a particular spin axis produce lower stretching exponents than static ones.

The above physical picture explains the easy-axis regime timescales, but not the easy plane. On the easy-axis side, the static Z fields lead to rapid, linear accumulation of precession phase, which causes rapid decay of the X and Y correlation functions. 
This prediction agrees with the $\lambda_{XXZ}<0$ region of Fig.~\ref{fig:fig3}(b), where the ratio of X to Z timescales (plotted in Fig.~\ref{fig:fig4}(e)) remains below 1. 
However on the easy plane side, and in particular for the XY point $H\propto S^x S^x+S^y S^y$, we would expect Z fields to decay faster than X and Y fields, and therefore the X to Z timescale ratio to significantly exceed 1, simply because there are more fields transverse to it. 
This runs contrary to observation, as can be clearly seen in the $\lambda_{XXZ}>0$ region of Fig.~\ref{fig:fig3}(b) and Fig.~\ref{fig:fig4}(e). 
Figure 4(e) also shows the expectation from the single-spin dynamical mean-field model, which shows sizable deviations from experimental observations.

To address these discrepancies, we consider a model including exact coherent interactions between strongly coupled pairs of spins within the dynamical mean field framework, as is depicted schematically in Fig.~\ref{fig:fig4}(d) and compared against data in Fig.~\ref{fig:fig4}(e).
This improvement can be understood by noting that local conservation of magnetization within these clusters  significantly extends the Z decay timescale relative to the transverse axes.
Thus, the reduction of the peak in timescale ratio near the XY Hamiltonian constitutes a signature of hybridization of clusters of spins contributing to the thermalization dynamics.
It is noteworthy that both experiment and exact diagonalization (Fig.~\ref{fig:fig4}(e)) give a lower timescale ratio than the pair-spin model, suggesting that the experiment witnesses quantum correlations that go beyond two-body clusters.

\outline{Discussion and Outlook}
\emph{Discussions.}---
These observations reveal the interplay between coherent hybridization and local dephasing as a concrete thermalization mechanism in a closed many-body system, and open up a wide range of opportunities for explorations in many-body physics and quantum sensing.
The disorder-order technique allows us to probe complex many-body phenomena where local control is otherwise inaccessible.
It will also be interesting to extend the Hamiltonian engineering techniques to higher spin dimensions, utilizing the full spin-1 nature of the NV center spin to access a richer range of dynamical phenomena~\cite{choi2017dynamical,schecter2019weak}, or to extend the measurement techniques for local autocorrelations to more intricate spin correlators, such as out-of-time-ordered-correlators~\cite{wei2018exploring,garttner2017measuring}.
Finally, our results highlight the ability to engineer complex interaction Hamiltonians and probe the resulting evolution, a key building block towards the use of such solid-state ensemble spin systems for entanglement-enhanced quantum sensing~\cite{davis2016approaching,hosten2016quantum,cappellaro2009quantum}. The techniques demonstrated here should be applicable  
to a wide variety of quantum simulation and sensing platforms.

\nocite{schachenmayer2015many,sachdev1993gapless,zhou2021disconnecting}


\emph{Acknowledgements.}---
We thank Yimu Bao, Paola Cappellaro, Joonhee Choi, Jordan Cotler, Eugene Demler, Wen Wei Ho, Francisco Machado, Daniel Parker, Pai Peng, Alex Schuckert, Norman Yao, Bingtian Ye, Bihui Zhu and Chong Zu for helpful conversations. We also thank Junichi Isoya, Shinobu Onoda, Hitoshi Sumiya and Fedor Jelezko for providing the black diamond sample. This work was supported in part by CUA, HQI, Vannevar Bush Faculty Fellowship Program, ARO MURI, DARPA DRINQS, Moore Foundation GBMF-4306.

Note added: Recently, we became aware of related work~\cite{peng2023exploiting}, which develops similar techniques and applies them to study spin transport.

\bibliography{main}

\end{document}


\title{Supplementary Materials for\\ ``Controlling Local Thermalization Dynamics in\\ a Floquet-Engineered Dipolar Ensemble"}

\author{Leigh S. Martin$^{1,\ast}$, Hengyun Zhou$^{1,\ast,\dagger}$, Nathaniel T. Leitao$^{1,\ast}$, \\Nishad Maskara$^1$,Oksana Makarova$^{1,2}$, Haoyang Gao$^1$, \\Qian-Ze Zhu$^{1,2}$, Mincheol Park$^1$,Matthew Tyler$^1$,\\ Hongkun Park$^{1,3}$, Soonwon Choi$^4$, Mikhail D. Lukin$^{1,\dagger}$\\
\normalsize{$^\ast$These authors contributed equally to this work.}\\
\normalsize{$^\dagger$To whom correspondence should be addressed;}\\
\normalsize{E-mail: hzhou@g.harvard.edu,  lukin@physics.harvard.edu}}

\maketitle
\tableofcontents

\pagebreak 
\section{Experimental Details}

\subsection{Sample and Experimental System}

Our spins are the $m_s=0$ and $m_s=-1$ levels of nitrogen vacancy center defects in diamond. The sample is a nanobeam of `black diamond,' the same sample used in Ref. \cite{choi2020robust}. The NV density is estimated to be $\sim 15$ ppm, which limits the XY8 decay timescale $T_2 = 1.6 \mu s$. Due to strain and the presence of other defects, $T_2^*$ is a relatively short 60 ns, which necessitates the use of advanced Hamiltonian engineering pulse sequences (see Sec. \ref{sec:HamEng}).

The sample is mounted on a stripline with an omega loop for driving microwave pulses. We illuminate the sample with a green laser (532 nm) and collect NV fluorescence for readout using a room temperature confocal microscope with an oil-immersion objective (Nikon 100x, NA=1.3). A dichroic mirror and 650 nm long-pass filter selectively passes the NV fluorescence, which is sent via fiber to a multi-pixel photon counter (Hamamatsu C14452-1550GA) with quantum efficiency of 25\%. The sample is mounted on a piezoelectric stage so that the focal spot may be stabilized to a desired point on the nanobeam. We apply pulse sequences using direct synthesis from a 12 gigasample per second arbitrary waveform generator (Tektronix model AWG7122C).

\subsection{Hamiltonian Engineering}
\label{sec:HamEng}

\begin{figure}[h!]
    \centering
    \includegraphics[width=\textwidth]{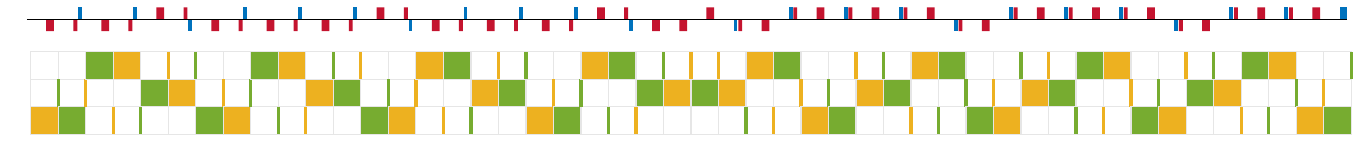}
    \caption{Illustration of the DROID-R2D2 sequence, see Ref.~\cite{tyler2023higher} for more details of the design procedure. The illustrations employ the same format as in Ref.~\cite{choi2020robust}. The top row shows the the pulses to be applied, with blue(red) pulses being around the X(Y) axis, and thick(thin) bars being $\pi$($\pi/2$) pulses. The bottom row shows the frame matrix representation, illustrating the structure in the pulse sequence.}
    \label{fig:R2D2}
\end{figure}

The native Hamiltonian in black diamond, when restricting to the $m_s=0$ and $m_s=-1$ sublevels of the NV center, is
\begin{align}
    H_{BD} &= H_0 + H_1 \\
    H_0 &= \sum_i h_i S_i^z \\
    H_1 &= \sum_{ij} J(\bm {r}_{ij}) \left(S_i^x S_j^x +S_i^y S_j^y - S_i^z S_j^z \right)
\end{align}
with $\overline{h_i}=0, \overline{h_i h_j} = \delta_{ij} W^2$, $W= 2\pi \left(4 \text{ MHz} \right)$ being the bandwidth of the normally-distributed disorder, 
$\bm r_{ij} = \bm r_i - \bm r_j$ is the interspin vector between NV centers $i, j$ that are coupled by the long-range dipolar potential
\begin{align}
J(\bm r_{ij})  = \frac{j_0 \, \left(3 \left(\hat{\bm B}_0 \cdot \hat{\bm r}_{ij}\right)^2 -1 \right)}{r^3_{ij}}, 
\end{align}
with anisotropy induced by the external quantizing field, $\hat{\bm B}_0= \hat{\bm z}$. 
The interaction scale at typical NV-NV separation, $a=11$ nm, is $(2\pi)(35 \text{ kHz})$.

We prolong the NV coherence time and engineer the native dipole-dipole Hamiltonian using the DROID-R2D2 (Disorder RObust Interaction Decoupling - Robust To Disorder at 2nd order) Floquet pulse sequence described in Ref.~\cite{zhou2023robust,tyler2023higher}, with the full pulse sequence given therein and also shown in Fig.~\ref{fig:R2D2}.
This sequence is a significant improvement over previous state-of-the-art sequences in Ref.~\cite{choi2020robust}: In addition to robustness against the leading order disorder and interaction contributions during the free evolution and finite duration pulses, we also systematically cancel certain higher-order terms in the Magnus expansion, systematically removing the leading artifacts in the DROID sequence for the task of Hamiltonian engineering, thus enabling higher fidelity to the target XXZ Hamiltonian.
To engineer the Heisenberg Hamiltonian (\textit{i.e.}, a decoupling sequence), all pulse spacings are set to 25 ns, with the exception of composite $\pi/2$ pulses, which consist of two $\pi/2$ pulses separated by a 1 ns delay to prevent pulse overlap.

As the pulse sequence consists only of $\pi/2$ and $\pi$ pulses about the cardinal axes $x,y,z$ of the Bloch sphere, the native interaction transforms among $XX+YY-ZZ$, $ZZ+XX-YY$ and $YY+ZZ-XX$, which we refer to as the $x$, $y$ and $z$ frames respectively.
To tune away from the Heisenberg point to an arbitrary XXZ Hamiltonian, we change the 25 ns pulse spacing in the $z$ frame to a new value between 2 and 70 ns.
The durations during the $x$ and $y$ frames are changed to keep the total Floquet period fixed to 1702 ns.
Note that although the native NV dipolar Hamiltonian is $S_xS_x+S_yS_y-2S_z S_z$, it becomes $XX+YY-ZZ$ when one restricts to the $m_s=0$ and $m_s=-1$ sublevels (where $X$, $Y$ and $Z$ are spin 1/2 angular momentum operators).
If the native Hamiltonian were $XX+YY-2ZZ$, then a decoupling sequence would simply cancel the Hamiltonian and XXZ Hamiltonian engineering would not be possible.

\subsection{Subsampling}
The disorder strength in our sample is roughly two orders of magnitude larger than the characteristic interaction strength. The DROID-R2D2 sequence is heavily optimized to cancel the resulting imperfections and higher-order terms. The result is a sequence with many pulses, and thus the Floquet period is not always short relative to the decay timescale under the engineered Hamiltonian. In order to measure the Ramsey and disorder order traces sufficiently many times before complete decay, we also perform measurements at fractional intervals of a Floquet cycle when the intrinsic decay is too fast to sample at integer multiples of the Floquet period alone. We choose these subsample points at 639 and 1053 ns, which minimizes residual disorder and interactions terms within average Hamiltonian theory.

\subsection{Tomography and Variance Subtraction}
Due to the beyond-rotating wave approximation effects described in Sec. \ref{Sec:FrequencyOpt}, DROID-R2D2 applies a small, unwanted rotation about the $z$ axis. In order to capture all residual polarization in our Ramsey and XY disorder order measurements, we measure along both the $x$ and $y$ axes and combine the results as $\mathcal{C} = \sqrt{\mathcal{C}_x^2 + \mathcal{C}_y^2}$. As noise in $\mathcal{C}_x$ and $\mathcal{C}_y$ will tend to increase $\mathcal{C}$, we subtract the noise estimated from error bars, \textit{i.e.}, $\mathcal{C} = \sqrt{\mathcal{C}_x^2 + \mathcal{C}_y^2 - (\delta \mathcal{C}_x^2 + \delta\mathcal{C}_y^2)/2}$. We do not attempt to compensate for unwanted $z$ rotations when measuring $\mathcal{C}_{DO}^{XZ}$ and $\mathcal{C}_{DO}^{ZY}$ disorder-order correlation functions, as $z$ rotations lead to imperfect rephasing and therefore cannot be recovered.

\subsection{Normalization and Fitting Methods}
\label{sec:normalization}
Ideally, a globally polarized state would not decay when engineering a Heisenberg Hamiltonian. However, higher-order imperfections and other effects inevitably limit coherence time in practice, as seen in Fig. 2a of the main text. Beyond pulse sequence design, we take several steps to minimize the impact of these imperfections on our data and stretched exponential fits. Firstly, to prevent $T_1$ decay and charge dynamics from contributing to the observed decay, we fix the time window between NV initialization into $m_s=0$ and readout, regardless of the number of Floquet cycles applied. Secondly, we normalize all of our data by the polarization decay observed at the Heisenberg point. For the disorder-order traces, we perform additional measurements to remove the effect of finite noise correlation time. This procedure, along with details of the disorder order pulse sequence, are detailed in Sec. \ref{Sec:SuppressDOInt}. Finally, we normalize the overall fluorescence contrast level by the level measured at $t=0$. To reduce the effect of slow drifts, the denominator of the normalization level is computed from all measurements performed in the 9 hours surrounding the measurement.

After normalization, we fit the data to stretched exponentials $\mathcal{C} = a e^{-(t/\tau)^\nu}$. We fix $a$ to equal the value of the first data point in each time series, and discard all time points after the first point with a contrast below 0.2. Changing this cutoff value does not significantly change the fitted $\tau$ or stretching exponent $\nu$, nor does treating $a$ as a fit parameter.

\subsection{Optimizing Microwave Frequency to Minimize Pulse Errors}
\label{Sec:FrequencyOpt}

\begin{figure}[h!]
    \centering
    \includegraphics[width=0.6\textwidth]{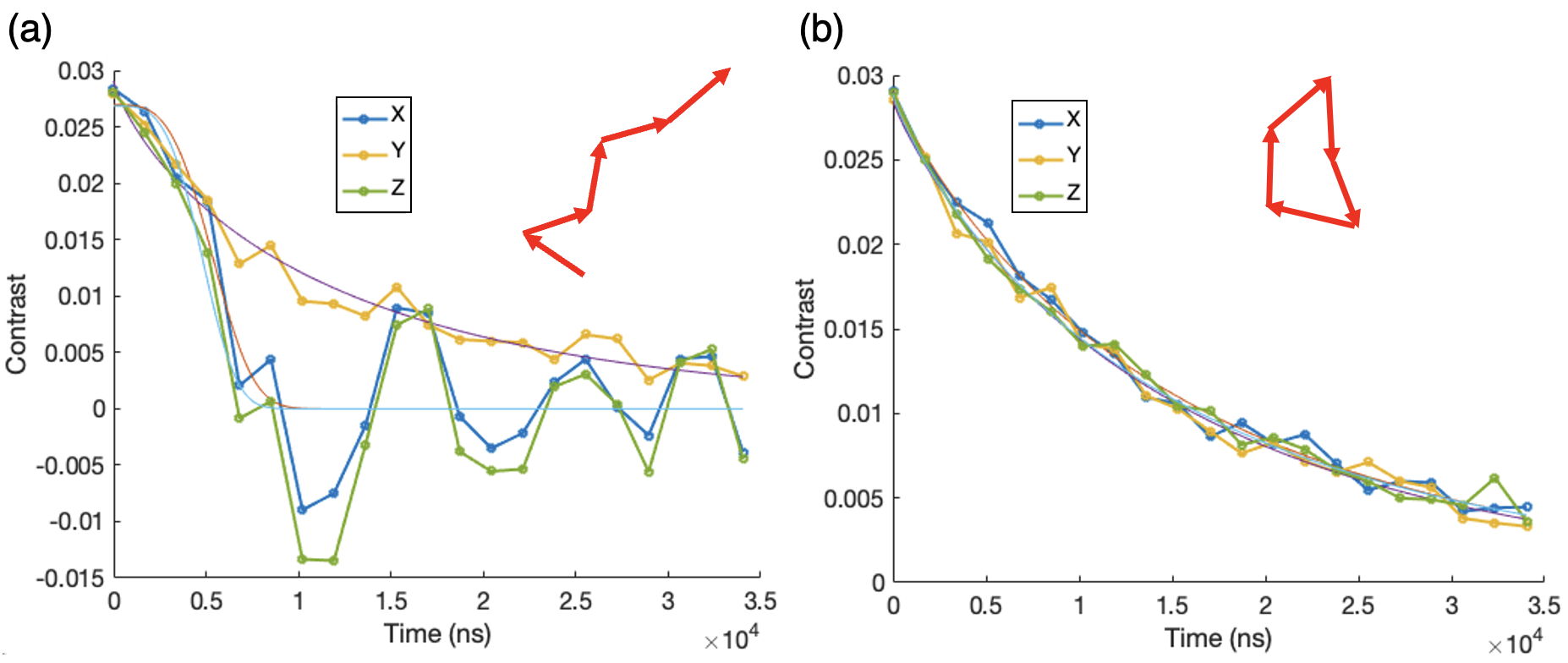}
    \caption{Decoupling behavior over time of spin system with different driving microwave frequencies: (a) Bad error case where systematic error occurs in spin control, showing oscillations with individual pulse errors compounding constructively; (b) Good error case where no systematic error occurs in spin control, showing clean decay with individual pulse errors compounding destructively.}
    \label{fig:MP-1}
\end{figure}

In order to control the NV spin system in our experiment, a sequence of AC microwave pulses is applied.
However, depending on the frequency of such microwave pulses, significant systematic errors can sometimes occur, as exemplified by the oscillations in Fig.~\ref{fig:MP-1}(a).
In comparison, a nearby frequency (1 MHz detuned, smaller than the 4 MHz inhomogeneous broadening of the spin system) shows a clean decay without oscillations (Fig.~\ref{fig:MP-1}(b)).
This suggests that microwave pulse errors are more prominent at some frequencies than others, causing global rotations of the spins at those frequencies.
In this section, we will characterize such pulse errors and determine frequencies that minimize these undesired rotations.
We find that the frequency-dependence is a result of different compounding behavior of individual pulse errors, which can be minimized by judicious choice of microwave drive frequency.

To understand these errors, we record the microwave waveform transmitted through the microwave stripline on a fast oscilloscope (20 GSa) and obtain the evolution unitary by numerically solving Schr\"{o}dinger's equation for a single spin in the presence of the measured control field.
The difference between the actual and ideal spin rotation can be expressed as an ``error angle'' along the $x$, $y$, $z$ axis, characterizing the magnitude of the systematic error that we wish to eliminate in our experiments.

\begin{figure}[h!]
    \centering
    \includegraphics[width = 0.6 \textwidth]{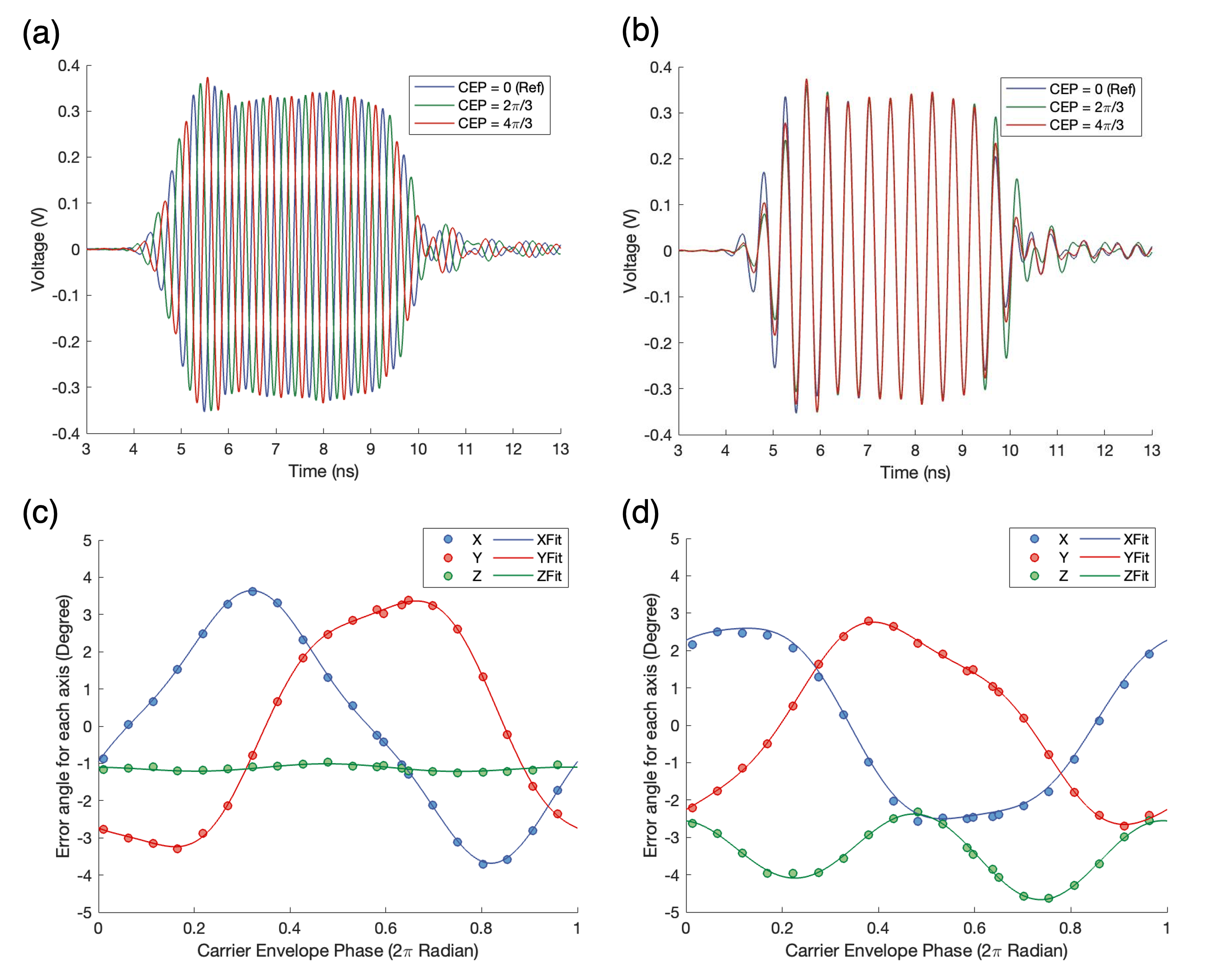}
    \caption{Overlapped waveform data vs. time (a) for different carrier envelope phase (CEP), showing different shapes and transients when time shifted (b). Error angles as a function of CEP for $\pi$ pulses (c) and composite $\pi/2$ pulse (d) at microwave frequency 2.0785 GHz. Note that each error angle's fitting function of CEP is a linear combination of $2\pi$, $\pi$, and $2\pi/3$ periodic sine functions with constant shift.}
    \label{fig:MP-2}
\end{figure}

In order to speed up the numerical computation and reduce waveform data sizes, we measure the waveforms of individual $\pi$ pulses and composite $\pi/2$ pulses at a number of microwave frequencies and compute the error angles, which we then use to obtain the total error of the full Floquet sequence.
Crucially, we obtain the error angle as a function of the carrier-envelope phase (CEP) \textit{i.e.}, the microwave phase at the start of the pulse envelope, as the pulse waveform transients can be modified by the CEP.
Because there are a limited number of cycles in the waveform, the rotating wave approximation may break down and pulse rotation can depend nontrivially on the CEP.
Also, since the waveform's envelope is not uniform during the pulse, the waveform shape varies with CEP, see Fig. ~\ref{fig:MP-2}(a) and (b).
We find that the CEP-dependent error angles are well-approximated by a linear combination of sinusoidal functions, see Fig.~\ref{fig:MP-2}(c) and (d).
In order to build up the correct Floquet unitary in the lab frame, we carefully account for time shifts to select the correct CEP for a given pulse.
Finally, we confirm that the Floquet unitary built from individual pulse building blocks is consistent with that obtained from a direct simulation of the complete waveform.

Examining the results for different microwave frequencies, we find that although individual pulse errors are comparable at different frequencies, the total error angle can be drastically different.
This suggests that the main source of different error behavior of the entire sequence, as shown in Fig.~\ref{fig:MP-1}, is the way individual errors add up; for ``bad" frequencies, the errors add up constructively due to the frequency choice, while for ``good" frequencies, the errors add up randomly and destructively interfere, see insets of Fig.~\ref{fig:MP-1}.

\begin{figure}[h!]
\centering
\includegraphics[width= 0.6\textwidth]{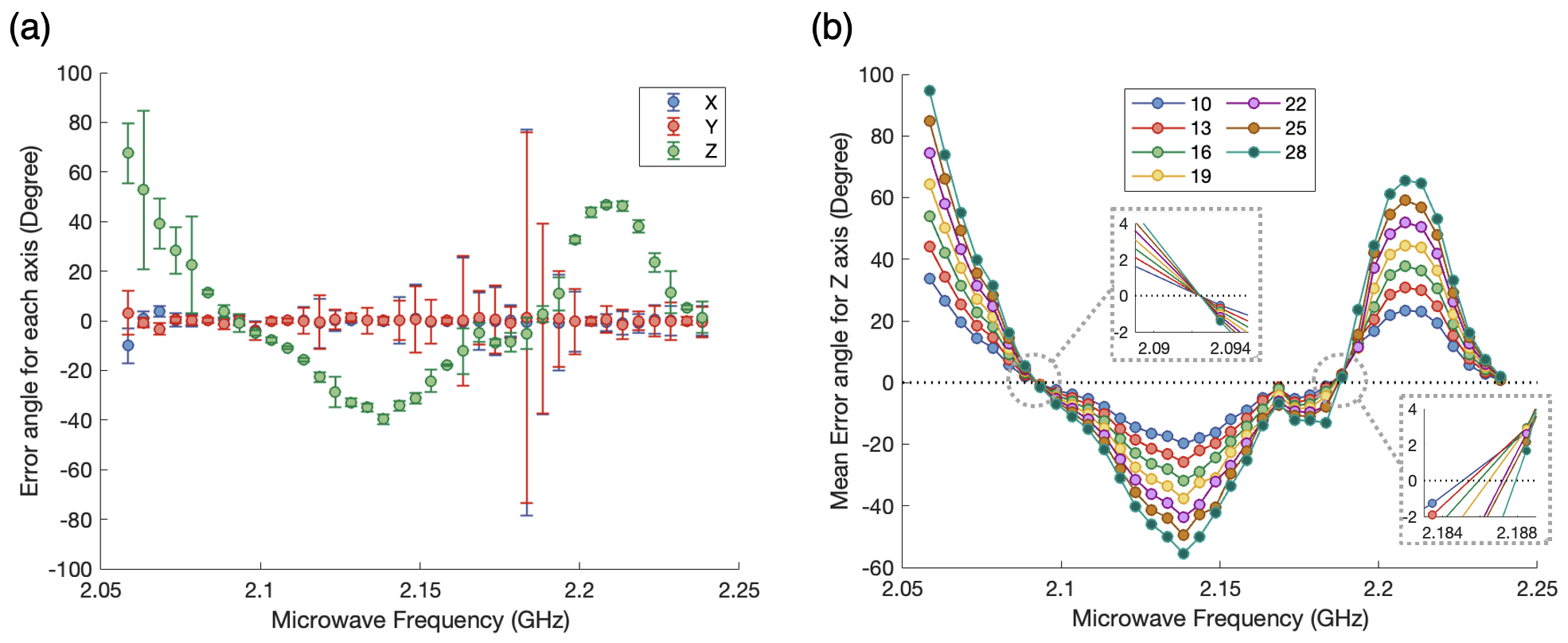}
\caption{Error angles as a function of microwave frequency (a) after 20 Floquet cycles and (b) a varying number of Floquet cycles of the DROID-R2D2 pulse sequence. The data point and error bar represent the average and the standard deviation of the error angle over different CEPs at a given frequency, respectively. Two ``zero-crossings'' are observed. The zero-crossings occur between 2.09 GHz and 2.094 GHz and between 2.184 GHz and 2.188 GHz. The second zero-crossing frequency varies significantly with the number of Floquet cycles, so we optimized the microwave frequency near the first zero-crossing.}
        \label{fig:MP-3}
\end{figure}

Using these simulations, we evaluate the error angles for a wide range of microwave frequencies and CEPs.
In Fig.~\ref{fig:MP-3}, we show the results, where the data points and error bars at each frequency indicate the mean and standard deviation of error angles over the CEP.
Varying the microwave frequency between 2.0585 GHz and 2.2385 GHz with 5 MHz sampling spacing, the average $z-$axis error angle after 20 cycles of the DROID-R2D2 pulse sequence shows a clear dependence on the microwave frequency, as shown in Fig.~\ref{fig:MP-3}(a) (note that the $x$ and $y$ axis mean error angles are much smaller, as the net $\pi$ rotation around $z$ every Floquet cycle echoes them out).
We observe two ``zero-crossings'' of the $z-$axis mean error angle occurring at $2.090-2.094$ GHz and $2.184-2.188$ GHz, as shown in Fig.~\ref{fig:MP-3}(b), regardless of the number of Floquet cycles used to control the spin system.
After taking more waveform data with finer microwave frequency spacing for these intervals, we find that 2.091872 GHz not only shows the smallest average and standard deviation of error angle over CEP, but also shows good robustness with small standard deviation of error angle over CEP within a considerable frequency range ($\pm 40$ kHz).

\begin{figure}[h!]
    \centering
    \includegraphics[width = 0.6\textwidth]{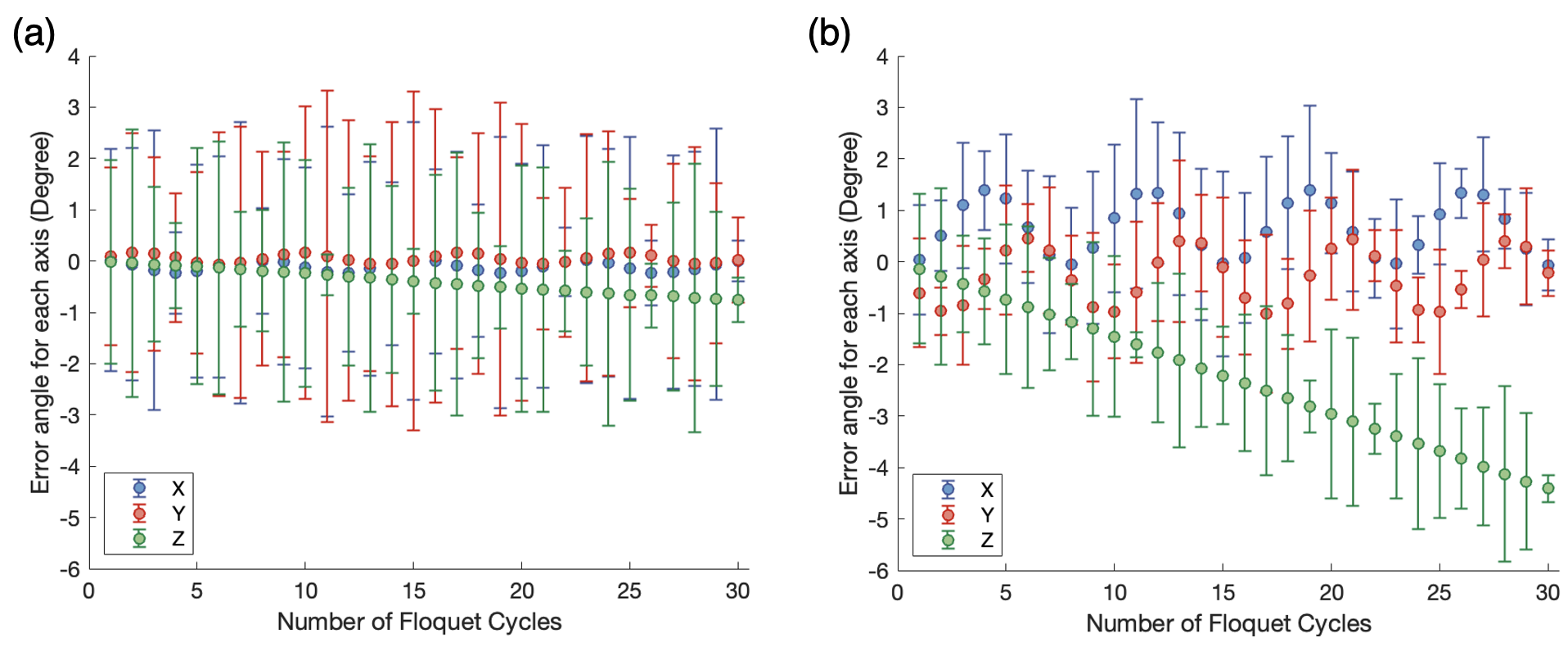}
        \caption{Error angles as a function of the number of Floquet cycles for the DROID-R2D2 pulse sequence with (a) $X = 16.1667$ ns and $Z = 42.6667$ ns and (b) $X = 25$ ns and $Z = 25$ ns. Both pulse sequences use the microwave frequency 2.091872 GHz. This full Floquet cycle compounding result confirms the frequency choice.}
        \label{fig:MP-4}
\end{figure}

In order to further confirm that 2.091872 GHz is a good frequency that minimizes spin control errors, we measure the waveform data of a full Floquet cycle of the DROID-R2D2 pulse sequence, extract the error angle vs. CEP via a simple periodic fitting function for the full Floquet cycle, and compound the error over different numbers of cycles.
We indeed find that this frequency produces only $\pm 6$ degree error in each axis, including the standard deviation shown in Fig.~\ref{fig:MP-4}, even with varying X and Z values for Hamiltonian engineering purposes.
This is further verified through direct measurements of the spin state under the sequence, where no prominent global rotations were observed.

\section{Disorder-Order Measurement}

In this section, we analyze the ability of our disorder-order measurement technique to use disorder as a resource to measure local operator correlation functions. 
We also consider imperfections induced by dynamic disorder and weak interactions, and describe our mitigation strategies. 

\subsection{Ideal Measurement}

The sequence depicted in Fig.~2(c) of the main text prepares the state
\begin{align}
    \ket{\psi_\theta(t)} &= R_\theta^\dagger U(t) R_\theta \ket{+}^{\otimes N}, 
\end{align}
where $R_{\theta} = \prod_j \exp{-i \theta_j S_j^z}\approx e^{-i H_{BD} \tau}$ is a product of local disordered rotations imprinting rotation angles $\theta_i = h_i \tau$, each of which is modelled as independently sampled from a zero mean Gaussian probability distribution of standard deviation $W\tau= (2\pi)(4$ MHz$)(0.2 \mu s)= 1.6 \pi$, where $\tau$ is the free-evolution time under the disorder field and $W$ is the intrinsic on-site disorder experienced by the spins in our sample. 
Measurement of the global $S^x$ magnetization yields the signal
\begin{align}
    \mathcal{S}(t) &= \overline{\bra{\psi_\theta(t)} \sum_i S_i^x \ket{\psi_\theta(t)}}  =  \prod_j \int_{-\infty}^{\infty} \frac{d\theta_j}{\sqrt{2\pi}W\tau} \exp{- \frac{\theta_j^2}{2 \left(W \tau\right)^2}} \bra{\psi_\theta(t)} \sum_i S_i^x \ket{\psi_\theta(t)}\\ 
    &= \sum_i \overline{ \text{Tr}\left\lbrack R_\theta^\dagger U(t) R_\theta \ket{+}^{\otimes N}\bra{+}R_\theta^\dagger U^\dagger(t)R_\theta  S_i^x \right\rbrack}, 
\end{align}
averaged over spatially uncorrelated local disorder. 
Using the invariance of the trace under cyclic permutations, we can rewrite this expression as a sum of local spin operators measured along the rotation axis $\hat{n}_{\theta_i} = \cos{\theta_i}\, \hat{x}+\sin{\theta_i}\, \hat{y}$ in a random product state locally correlated with the measurement axis,

\begin{align}
    \mathcal{S}(t) &= \sum_i \overline{ \text{Tr}\left\lbrack \left(R_\theta \ket{+}^{\otimes N}\bra{+}R_\theta^\dagger \right) \left(U^\dagger(t)R_\theta  S_i^x R_\theta^\dagger U(t) \right) \right\rbrack} \\
    &= \sum_i \overline{ \text{Tr}\left \lbrack  \prod_j \frac{1+ \hat{n}_{\theta_j} \cdot \sigma_j }{2} \hat{n}_{\theta_i} \cdot \bm S_i(t)\right \rbrack }.
\end{align}
Physically, when the spins are fully depolarized and wound by the disorder to be evenly spaced around the XY plane ($W\tau \gtrsim \pi$), the local mean of the spin texture $\overline{n^\mu_\theta}$ vanishes but covariance $\overline{n_\theta^\mu n_\theta^\nu}$ projects onto the $XY$ plane, relative to errors exponentially small in $(W\tau)^2$. 
This ensures that only the \textit{local autocorrelations} of operators defined on this plane contribute to the experimental signal yielding 
\begin{align}
    \mathcal{S}(t) &= \sum_i \overline{\hat{n}^\mu_{\theta_i}\hat{n}^\nu_{\theta_i}}  \frac{1}{2^{L-1}}\text{Tr}\lbrack S_i^\mu(t) S_i^\nu \rbrack + \mathcal{O}(e^{-(W \tau)^2})\\
    &= \sum_i \frac{\text{Tr}\left\lbrack S_i^x(t) S_i^x \right \rbrack }{\text{Tr}\left\lbrack 1 \right \rbrack}+\frac{\text{Tr}\left\lbrack S_i^y(t) S_i^y \right \rbrack }{\text{Tr}\left\lbrack 1 \right \rbrack} + \mathcal{O}(e^{-(W \tau)^2}) \\
    \mathcal{S}(t) &\equiv N\mathcal{C}_{DO}^{XY}(t)/2 + \mathcal{O}(e^{-(W \tau)^2}).
\end{align}

Note that we have fixed the overall normalization of the correlation function $\mathcal{C}_{DO}^{XY}$ so that it is at most unity. 
By applying $\pi/2$ pulses globally to the spin ensemble, effectively changing the plane of the spin texture, we can alternatively measure the correlators
\begin{align}
    \mathcal{C}_{DO}^{YZ}(t) &= \frac{2}{N}\sum_i \langle S_i^y(t) S_i^y(0) \rangle_{T=\infty} + \langle S_i^z(t)S_i^z(0) \rangle_{T= \infty},\nonumber  \\
    \mathcal{C}_{DO}^{ZX}(t) &= \frac{2}{N}\sum_i  \langle S_i^z(t)S_i^z(0) \rangle_{T= \infty}+\langle S_i^x(t) S_i^x(0) \rangle_{T=\infty}.
\end{align}
Three independent measurements of initial states wound around the $\lbrace XY, YZ, ZX \rbrace$ planes therefore allow us to infer the autocorrelations of three independent spin components
\begin{align}
    \mathcal{C}^{\mu\mu}_{\text{Local}}(t) = \frac{4}{N}\sum_i \langle S_i^\mu(t) S_i^\mu(0) \rangle_{T=\infty},
\end{align}
through linear recombination, 
\begin{align}
    \mathcal{C}_{\text{Local}}^{XX} &=  +\mathcal{C}_{DO}^{XY} - \mathcal{C}_{DO}^{YZ} + \mathcal{C}_{DO}^{ZX}, \\
    \mathcal{C}_{\text{Local}}^{YY} &=  +\mathcal{C}_{DO}^{XY} + \mathcal{C}_{DO}^{YZ} - \mathcal{C}_{DO}^{ZX}, \\
    \mathcal{C}_{\text{Local}}^{ZZ} &= - \mathcal{C}_{DO}^{XY} + \mathcal{C}_{DO}^{YZ} + \mathcal{C}_{DO}^{ZX}.
\end{align}

\subsection{Effects of Dynamic Disorder}
In order to extract the thermalization dynamics caused purely by many-body interactions from the measured decay curves, we need to normalize the measured decay curves by contributions from other decay sources.

For measurements of global auto-correlators via Ramsey sequences, the normalization procedure in Sec.~\ref{sec:normalization}, where we normalize by the decay of a polarized initial state under engineered Heisenberg interactions, captures the dominant Hamiltonian engineering imperfections and other sources of decay.
For measurements of local autocorrelators via disorder-order sequences, there is however an additional source of decay of non-ideal unwinding, coming from imperfect time-correlations of on-site disorder during the winding and unwinding steps.

\begin{figure}[h!]
    \centering
    \includegraphics[width = 0.4\textwidth]{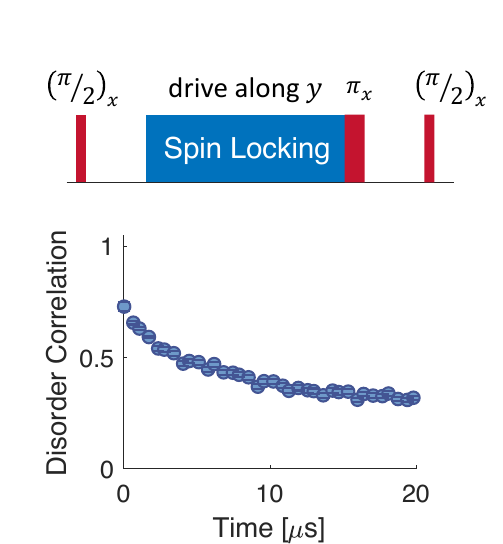}
    \caption{\textbf{Dynamical disorder normalization.} (a) Diagram of the experiment we do to normalize out contributions from dynamical disorder. The experiment contains three steps: winding under disorder, spin locking to freeze the dynamics, and unwinding under slightly changed disorder after some time. (b) The measured contribution due to dynamical disorder (i.e. $\frac{2S_{DO+SL}\left(t\right)}{S_{SL}\left(t\right)}$ in Eq.~(\ref{eq:normalization_formula})).}
    \label{fig:dynamic_disorder}
\end{figure}

The contribution of dynamical disorder can be measured by a spin locking experiment, as shown in Fig.~\ref{fig:dynamic_disorder}. In this experiment, we perform the same disorder winding and unwinding step as our disorder-order measurement, but instead of engineering an XXZ Hamiltonian in the middle, we apply a spin locking sequence (i.e. a continuous rotation around the $y$ axis). The spin locking transforms the native Hamiltonian $S^xS^x+S^yS^y-S^zS^z$ into $S^yS^y$, which conserves the $y$ component of individual spins and therefore freezes the dynamics. Therefore, by measuring the signal after winding and unwinding steps separated by some time, we can extract the amount of signal decay originating from imperfect correlations of disorder during these two steps.

More specifically, if we denote the phase accumulated due to disorder during the winding and unwinding steps on spin $i$ by $\theta_1^i$ and $\theta_2^i$, the dynamics is the following: 
\begin{itemize}
    \item During the winding step, spin $i$ rotated by $\theta_1^i$ in XY-plane, leaving $y$ polarization $\cos{\theta_1^i}$.
    \item During the spin locking step, the $x$ component of the spins decay out under the $S^yS^y$ interaction, assuming that the spin locking time is much longer than $T_2$, and conserves the $y$ component $S_i^y=\cos{\theta_1^i}$. Even if the spin locking time is not much longer than $T_2$, we can still remove the $x$ component by averaging the final measured signal within a spin locking duration window $\Delta T = \frac{2\pi}{\Omega}$, where $\Omega$ is the Rabi frequency of the spin locking.
    \item During the unwinding step, spin $i$ is rotated by $\theta_2^i$ in XY-plane, resulting in a final $y$ component $\cos{\theta_1^i}\cos{\theta_2^i}$.
\end{itemize}
Therefore, the final measured signal is:
\begin{equation}
    S_{DO+SL}\left(t\right) = \overline{\cos{\theta_1^i}\cos{\theta_2^i}} = \frac{1}{2}\overline{\cos\left(\theta_1^i-\theta_2^i\right)}+\frac{1}{2}\overline{\cos\left(\theta_1^i+\theta_2^i\right)}.
\label{eq:dynamical_disorder_signal}
\end{equation}
The first term is the contribution of imperfect unwinding in the disorder-order experiment, and the second term is negligible as long as the winding and unwinding time is much longer than $T_2^*$, which is the case in our experiments. Therefore, the dynamical disorder normalization is done by dividing the measured decay curve by $2S_{DO+SL}\left(t\right)$.

Another detail of the dynamical disorder normalization is that the measurement of $S_{DO+SL}\left(t\right)$ is itself subject to decay during the spin locking, so it may not reflection the contribution of dynamical disorder faithfully. In order to overcome this, we normalized $S_{DO+SL}\left(t\right)$ itself by the same experiment without winding and unwinding steps (i.e. only spin locking with the same time duration). The final formula we use to normalize the data is:
\begin{equation}
    S_{Normalized}\left(t\right) = \frac{S_{DO+XXZ}\left(t\right)}{S_{Heisenberg}\left(t\right)\frac{2S_{DO+SL}\left(t\right)}{S_{SL}\left(t\right)}} = \frac{S_{DO+XXZ}\left(t\right)S_{SL}\left(t\right)}{2S_{Heisenberg}\left(t\right)S_{DO+SL}\left(t\right)},
\label{eq:normalization_formula}
\end{equation}
where $S_{DO+XXZ}\left(t\right)$ is the decay curve measured in the disorder-order experiment, $S_{Heisenberg}\left(t\right)$ is the decay curve of a polarized initial state under engineered Heisenberg interaction, $S_{DO+SL}\left(t\right)$ is the decay curved measured in the ``winding-spin locking-unwinding" experiment shown in Fig.~\ref{fig:dynamic_disorder}, and $S_{SL}\left(t\right)$ is the decay curve of a polarized initial state under spin locking.

\subsection{Interaction-Induced Imperfections and Suppression Methods}
\label{Sec:SuppressDOInt}

A main imperfection in the disorder-order measurement comes from dipole-dipole interactions during the winding and unwinding process. To understand the effects of this imperfection, let us go into the interaction picture with respect to the on-site disorder $h_i$. In this interaction picture, the spin operators are transformed into:
\begin{align}
    S_i^x&\rightarrow \cos{h_i t}S_i^x + \sin{h_i t}S_i^y,\nonumber\\
    S_i^y&\rightarrow \cos{h_i t}S_i^y - \sin{h_i t}S_i^x,\nonumber\\
    S_i^z&\rightarrow S_i^z.
\end{align}
Therefore, the native interaction $H_{int} = J_{ij}\left(S_i^x S_j^x + S_i^y S_j^y - S_i^z S_j^z\right)$ is transformed into:
\begin{align}
    \tilde{H}_{int} &= J_{ij}\left[\left(\cos{h_i t}S_i^x + \sin{h_i t}S_i^y\right)\left(\cos{h_j t}S_j^x + \sin{h_j t}S_j^y\right)\right.\nonumber\\&\left.+ \left(\cos{h_i t}S_i^y - \sin{h_i t}S_i^x\right)\left(\cos{h_j t}S_j^y - \sin{h_j t}S_j^x\right) - S_i^z S_j^z\right]\nonumber\\
    &= J_{ij}\left[\cos{\left[\left(h_i - h_j\right)t\right]}\left(S_i^x S_j^x + S_i^y S_j^y\right) + \sin{\left[\left(h_i - h_j\right)t\right]}\left(S_i^y S_j^x - S_i^x S_j^y\right) - S_i^z S_j^z\right].
\label{eq:H_transformed_by_disorder}
\end{align}
In our experimental system, on-site disorder is much stronger than the interactions. Therefore, for most pairs $\{i,j\}$, the difference of disorder $h_i-h_j$ is much larger than the interaction strength $J_{ij}$. For this reason, the first two terms in Eq.~(\ref{eq:H_transformed_by_disorder}) are rapidly averaged out, and the effective Hamiltonian during the winding and unwinding process is
\begin{equation}
    H_{eff} = -J_{ij}S_i^z S_j^z.
\label{eq:minus_Ising_during_winding}
\end{equation}

The Ising-like effective Hamiltonian Eq.~(\ref{eq:minus_Ising_during_winding}) pollutes our measurement of stretching exponent. As a specific example, consider the case where we measure the quenched dynamics at the Ising point $H_{int} = J_{ij}S_i^z S_j^z$. In this case, the effective Hamiltonian Eq.~(\ref{eq:minus_Ising_during_winding}) during the winding and unwinding steps can be viewed as an evolution backward in time. This causes the measured decay curve to decay very slowly or even increase a little bit at early time, which pushes the fitted stretching exponent up.

To overcome this systematic error, we concatenate the winding and unwinding steps with an additional Hamiltonian engineering step, where we engineer the Ising Hamiltonian $H_{int} = J_{ij}S_i^z S_j^z$ for the same time duration as the winding/unwinding steps. The Ising Hamiltonian is engineered by a continuous driving along $X$ axis, which engineers the Hamiltonian $J_{ij}S_i^x S_j^x$. This effective Hamiltonian is then transformed to $H_{int} = J_{ij}S_i^z S_j^z$ by a $\pi/2$ pulse that rotates the $X$ axis into the $Z$ axis.
In addition, we fine-adjust the separation between pulses in order to compensate for residual disorder effects during the sequence.
With these techniques, we have cancelled the undesired effective Hamiltonian Eq.~(\ref{eq:minus_Ising_during_winding}) to leading order, and removed an important artifact in the stretching exponent data.

\section{Numerical Simulations of Ideal Hamiltonian Engineering}
To gain a quantitative understanding of the fidelity of the Hamiltonian engineering achieved in the experiment, we proceed to benchmark our results against numerical simulations of the global and local autocorrelations generated by ideal XXZ dynamics with quench positional disorder. Specifically, we utilize Krylov time-evolution methods to exactly simulate the dynamics of small system sizes of up to $N=18$ spins, which are complemented by semi-classical discrete truncated Wigner approximation (dTWA) simulations of the system at large scale, $N = 200$.

\subsection{Methods}

\subsubsection{Simulation Procedures}

%
We proceed to describe the numerical methods utilized to simulate ideal Hamiltonian engineering associated to the dipolar XXZ Hamiltonian
\begin{align}\label{eq:XXZ-Dipoles}
    H_{\bm g} &= \sum_{ij} J(\bm r_{ij}) \,\sum_\mu  g_\mu S_i^\mu S_j^\mu.
\end{align}
In all our simulations, quench positional disorder is incorporated by sampling every dipole's position uniformly in a 3-dimensional system with periodic boundary conditions, of linear length $l=N^{1/3}a$, where $a = 11 $nm is the typical inter-spin separation of a single group of NVs in the experiment. 
Timescales are then expressed in units of $J =  j_0/a^3 = (2\pi)\left(35\, \text{kHz} \right)$, set by the characteristic interspin distance. 
For reference, the exact decay time of the Ising point in these units reads~\cite{feldman1996configurational}
\begin{align}
J \tau_{\text{Ising}} &= \frac{9 \sqrt{3}}{8 \pi^2} \approx 0.2,
\end{align}
which is verified in both numerical methods. 
We average over $M=100$ positional configurations in all our calculations henceforth.

To simulate the infinite temperature quench with Krylov evolution, we average over dynamics initialized in Haar-random states in the $2^N$ dimensional Hilbert space. Since the variance of any local observable over the Haar distribution vanishes exponentially in system size (quantum typicality), we find that a single sample is sufficient to characterize the infinite temperature local autocorrelation. 
To scale to system sizes as large as $N=18$ spins, we take advantage of the $U(1)$ symmetry of the dynamics and apply Krylov subspace approximations to the propagator in each charge sector independently, summing the results to obtain the appropriate correlation function.  
More explicitly, given a Haar random state $\ket{\psi}$, $3N$ additional states $\ket{\phi_i^\mu} = \sigma_i^\mu \ket{\psi}$ are prepared and the collection of $3N+1$ states are propagated by time-steps $\delta t = J/100$ in parallel over the respective charge sectors of the XXZ Hamiltonian. 
Local correlation functions are calculated as 
\begin{align}
C_{\text{Local}}^{\mu \mu}(t) = \frac{1}{N}\sum_i \frac{\text{Tr}\left(U^\dagger(t) \sigma_i^\mu U(t) \sigma^\mu_i \right)}{\text{Tr}(1)} \approx  \frac{1}{N}\sum_{i} \bra{\psi(t)} \sigma_i^\mu \ket{\phi_i^\mu(t)}.
\end{align}
%

To simulate the quench dynamics in dTWA, we sample over $n_t = 2000$ fluctuations of initial classical dipole configurations, characterized by a uniform average over Wooter's vector ensembles 
\begin{align}
\mathcal{E}_{\text{Ramsey}} &= \left \lbrace (1,1,1),(1,-1,-1), (1,1,-1),(1,-1,1) \right \rbrace \\
\mathcal{E}_{T=\infty} &= \left \lbrace (1,1,1),(1,-1,-1), (-1,1,-1),(-1,-1,1) \right \rbrace \\
\end{align}
which either describe a dipole polarized along $+\hat{x}$ or at infinite effective temperature, for global and local autocorrelation measurements respectively. 
In each instance of these fluctuations, we calculate the autocorrelation of each axis of the dipoles as they are numerically integrated by their classical equations of motion~\cite{schachenmayer2015many}. 
The average of the spin autocorrelations over these fluctuations of initial conditions are then taken as a semi-classical benchmark for the dynamics in the experiment.
\subsubsection{Fitting Procedure}

\begin{figure}[h!]
\centering
\includegraphics[width=\textwidth]{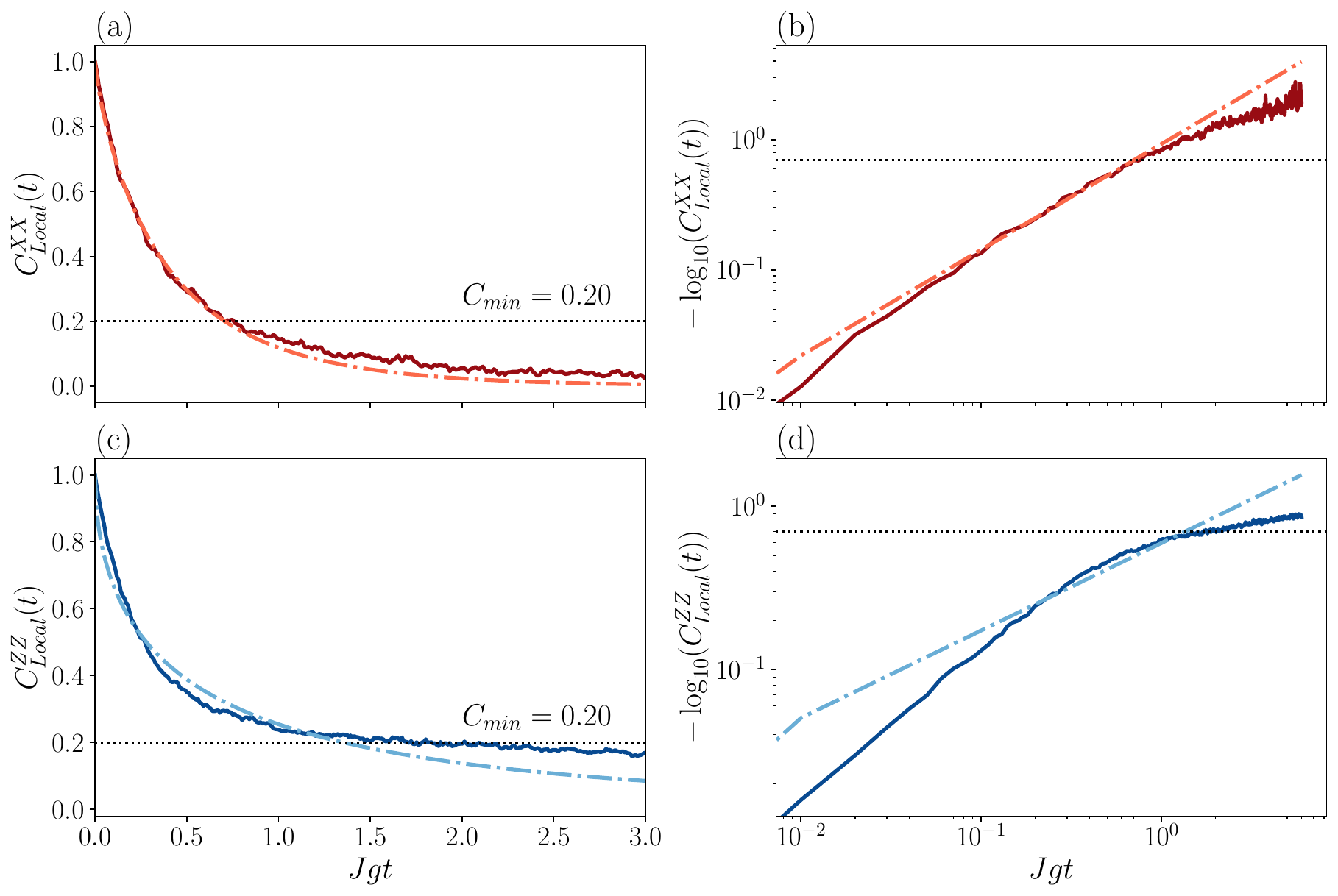}
\caption{\textbf{Stretched exponential fitting}: Characteristic local autocorrelation traces for XY interactions simulated with $L=18$ Krylov on (a,c) linear  and (b,d) triple logarithmic plot. First row depict $X$ autocorrelations and the last row dipict $Z$ autocorrelations, with solid lines arising from Kyrlov simulations and dashed lines the associated fit.}
\label{fig:fitting_dtwa}
\end{figure}

%
After an autocorrelation function has been calculated via either numerical procedure, we fit the corresponding decay trace to a stretched exponential while the signal is above a threshold of $C_{min} = 0.2$, which is a cut-off determined by the noise floor of the experimental data. 
An example fit is provided in Fig. \ref{fig:fitting_dtwa}, both on a linear plot as well as a triple logarithmic plot upon which exact stretched exponential decay appears as a straight line.

\begin{figure}[h!]
\centering
\includegraphics[width=\textwidth]{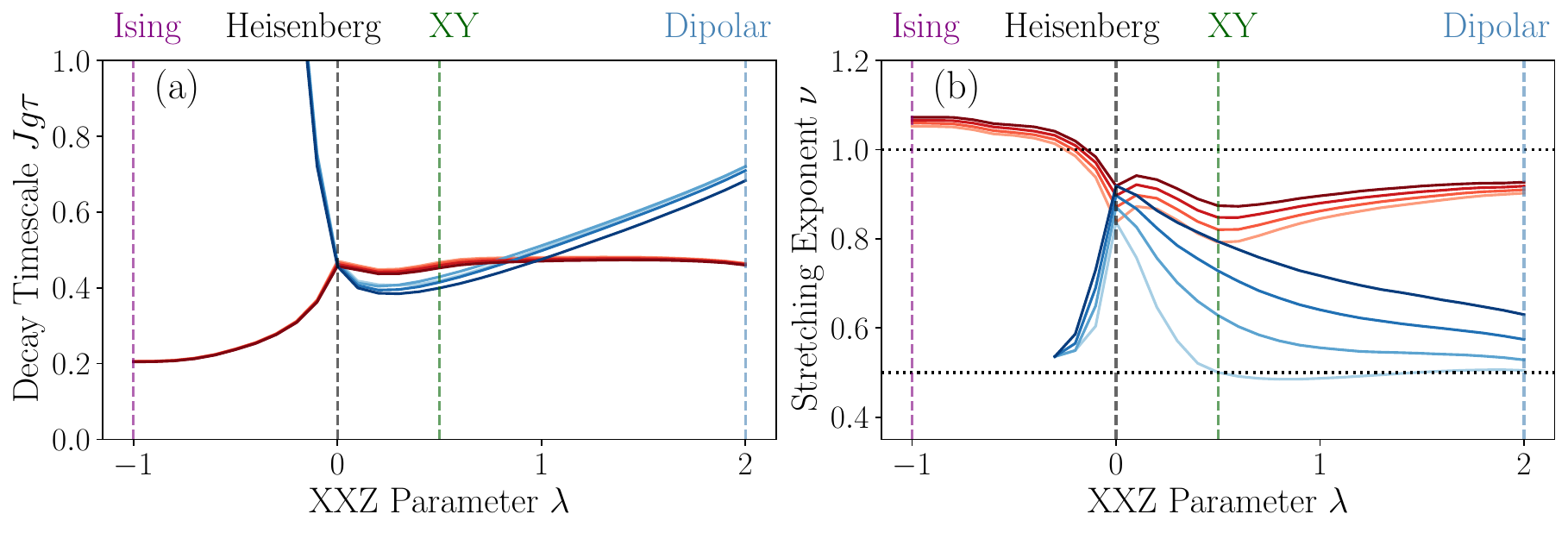}
\caption{\textbf{Flow of stretched exponential with threshold}: Stretched exponential fit parameters obtained from dTWA for different values of the fitting threshold. In order of increasing hue darkness for both X (red) and Z (blue) autocorrelations, the threshold varies from $C_{min} =  0.05,0.1,0.15,0.2$. }
\label{fig:fit_thres_flow}
\end{figure}
    
%
The essential \textit{qualitative} features of $X$ decay timescales and shapes are not modified by using a different value of this cut-off, as is explicitly checked in Fig.~\ref{fig:fit_thres_flow}.
The stretching exponent for the $Z$ decay, however, is in fact strongly sensitive to the fitting thresold particularly on the easy-plane side of the phase diagram, as is shown in Fig.~\ref{fig:fit_thres_flow}(b). 
This justifies why $Z$ stretching exponents were not included in the main text, as it is not a well-defined characterization of the relaxation dynamics; different time-windows admit different local stretching exponents, as is explicitly visible on the triple logarithmic plot of Fig.~\ref{fig:fitting_dtwa}(d).

\subsection{Global vs. Local Relaxation Timescales} \label{sec:global}

\begin{figure}[h!]
\centering
\includegraphics[width=\textwidth]{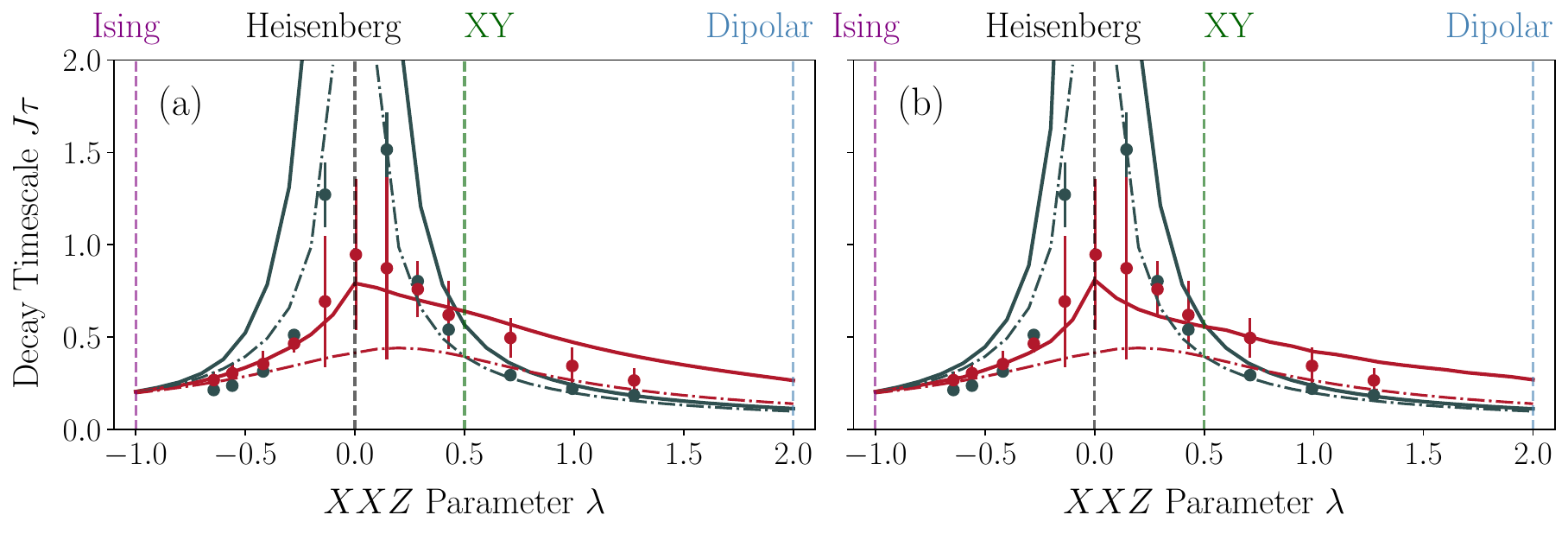}
\caption{\textbf{Benchmarking global and local $X$ decay}: Experimental Ramsey data (grey circles) compared to (a) $N=200$ dTWA  and (b) $N=18$ Krylov dynamics, both of which are plotted as solid lines. Local correlator decay is also plotted in red both from experiment and simualtion, rescaled by the factors defined in Sec. \ref{sec:local-benchmark}. Finally the dashed lines in both figures plot the analytical early time expansion prediction in Eq.~\ref{eq:earlytime_global_vs_local} which succesfully reproduces the cross-over of local and global timescales at the XY Hamiltonian. } 
\label{fig:global}
\end{figure}

%
We proceed to explicitly benchmark the comparison of local and global $X$ decay timescales reported in Fig.~2(b) in the main text. 
The main notable features from the experiment, including the divergence of the Ramsey timescale and cross-over between local and global timescales at the XY interaction Hamiltonian are both reproduced in numerical simulations. 
The first feature is already understood simply via symmetry principles explained in the main text. 
The latter can be explained by a perturbative early-time analysis, whose prediction is plotted in Fig.~\ref{fig:global} as a dashed line. 

%
In particular, one can calculate in time-dependent perturbation theory, expanding $O(t)=\exp{i\lbrack H,  \rbrack t}O$ for relevant operators $O$, 
\begin{align}
C^{XX}_{\text{G/L}}(t) = 1 - (\Gamma_{\text{G/L}} t)^2/2 + \mathcal{O}((Jt)^4)
\end{align}
where
\begin{align}\label{eq:earlytime_global_vs_local}
\Gamma^2_{\text{G/L}} &= \sum_i \begin{cases} 
\langle \lbrack H, \lbrack H , S_i^x \rbrack \rbrack \rangle_{+\hat{x}}  \qquad \text{Global}\\
\langle \lbrack H, \lbrack H , S_i^x \rbrack \rbrack S_i^x\rangle_{T=\infty} \qquad \text{Local}
\end{cases}
\\
&= \frac{1}{4}\overline{\sum_{ij} J^2_{ij}} \begin{cases} 
\left(g_x-g_z\right)^2  \qquad \text{Global}\\
g_{x}^2 + g_z^2 \qquad \text{Local}
\end{cases}
\end{align}
are the associated decay rates.
Physically, this cross-over can be understood by examining a pair of strongly coupled spins, and analyzing the relevant interaction energy scales between different states, resulting in different phase accumulation behavior.
In particular, for symmetric initial states prepared in Ramsey measurements, the singlet state does not participate in the dynamics, so the only relevant energy scale is the gap from the $\ket{00},\ket{11}$ states to the triplet state $\ket{T}= (\ket{01}+\ket{10})/\sqrt{2}$. 
On the other hand, at infinite temperature, all four eigenstates of the strongly coupled pair are relevant, including the gap to the previously dark singlet $\ket{S}= (\ket{01}-\ket{10})/\sqrt{2}$.
As we cross the XY point from left to right in Fig.~\ref{fig:global}, the 00/11-triplet gap starts to exceed the 00/11-singlet gap, resulting in the relative ordering of Ramsey decay time and infinite temperature decay time reversing. 
\subsection{Local Decay Timescales and Shapes}\label{sec:local-benchmark}
%
%
Simulations of local decay timescales and shapes also share qualitative agreement with the experimental data. 
However, we find that an overall timescale rescaling is necessary to obtain quantitative agreement between our simulations and the experimental data. 
The optimal rescaling values of $r_{\text{Krylov}} = 0.24$ and $r_{\text{dTWA}}=0.25$ are found by minimizing the following cost function, determined by experimental data/errors $\lbrace (\tau_\alpha, \Delta \tau_\alpha) \rbrace_\alpha$ and simulation data/errors $\lbrace (\tau'_\alpha, \Delta \tau'_\alpha) \rbrace_\alpha$ indexed by $\alpha$ encoding both the XXZ value and spin axis taking $\mathcal{N}$ values,
\begin{align}
C(r) = \frac{1}{2\mathcal{N}} \sum_\alpha \frac{\left(r \tau_\alpha-\tau'_\alpha \right)^2}{r^2 \left(\Delta \tau_\alpha \right)^2+(\Delta \tau'_\alpha)^2}.
\end{align}
\begin{figure}[h!]
\centering
\includegraphics[width=\textwidth]{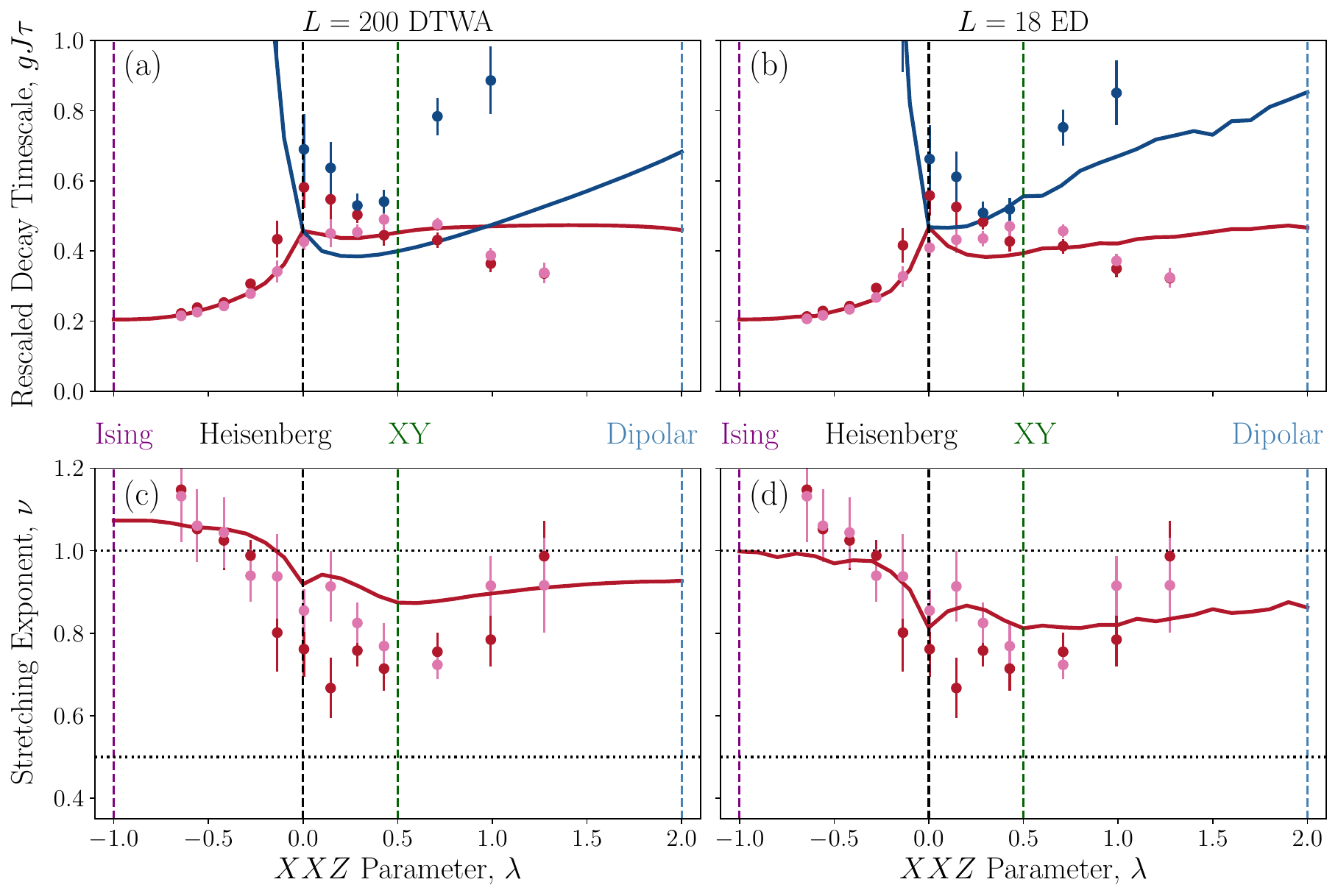}
\caption{\textbf{Benchmarking Hamiltonian engineering by simulation of local autocorrelations}: Fit parameters from experimental data for X (red circles), Y (pink circles) and Z (blue circles) local autocorrelation measurements compared against dTWA and Krylov subspace dynamics (lines). Global factors of $r = 0.25, 0.24$ have been applied to the experimental timescales to agree with the numerical calculations in the respective methods of dTWA and Krylov via the least squares fitting procedure described in the main text.}
\label{fig:local-benchmarks}
\end{figure}

We remark that the dTWA simulation shown here incorporates a UV cut-off, minimum interspin distance $r_{min}=0.2a$, unlike the Krylov simulation. This is the reason for the stretching exponent going above $d/\alpha=1$ in Fig. \ref{fig:local-benchmarks}(c), suggesting that the disagreement of the experimentally observed stretching exponent from the analytical Ising expectation is due to the minimal distance between NV centers associated to the diamond lattice.

\begin{figure}[h!]
\centering
\includegraphics[width=0.9\textwidth]{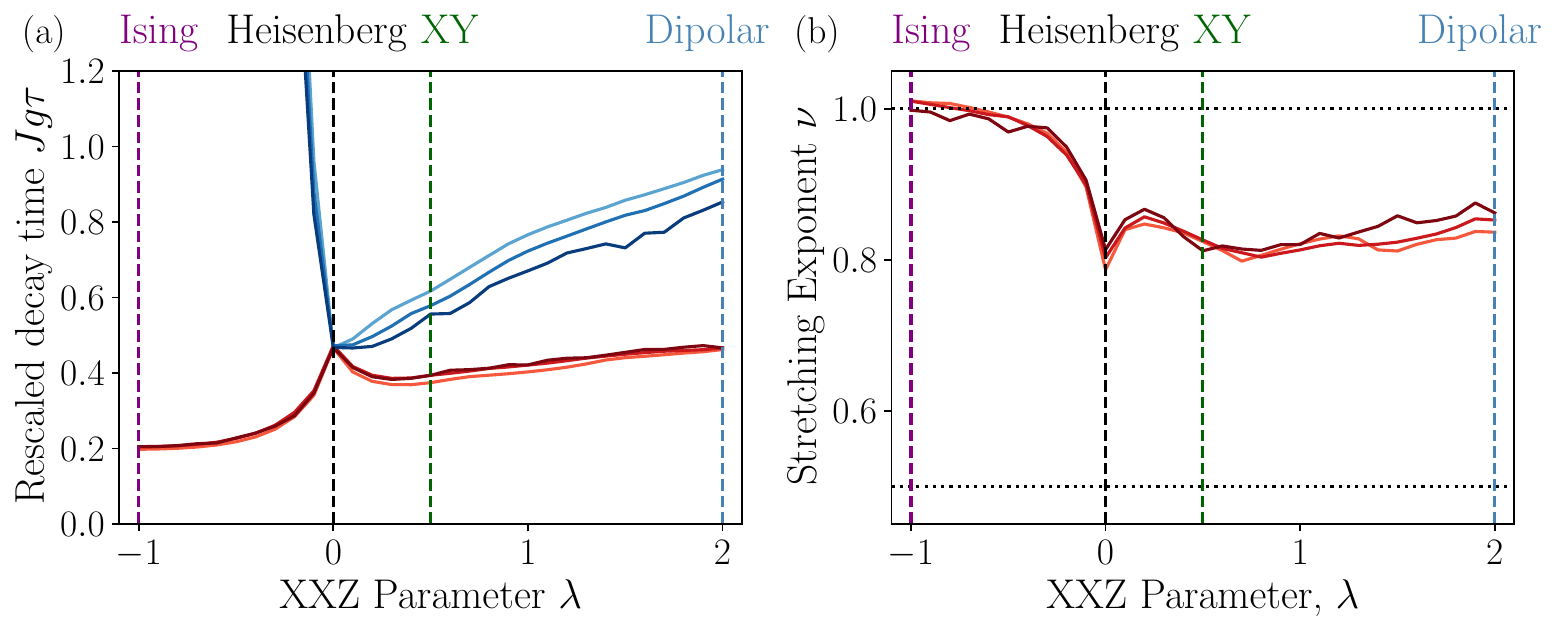}
\caption{\textbf{Finite size scaling of Krylov fits}: Stretched exponential fit parameters obtained from Krylov dynamics for different system sizes. In order of increasing hue darkness for both X (red) and Z (blue) autocorrelations, the system size increases as $N =  14,16,18$. While the $X$ decay is hardly modified, the flow of $Z$ decay parameters is more pronounced.}
\label{fig:fss-krylov}
\end{figure}

As a final verification of the results, we check that our exact Krylov simulations do not exhibit strong finite size effects that might invalidate the features in the stretching exponents and timescales. 
As is shown in Fig. \ref{fig:fss-krylov}, as the system size increases, the stretched exponential parameters of the X relaxation are essentially invariant. 
The Z timescales are found to flow to smaller values, albeit strictly larger than X in all simulations.

\subsection{Absence of Hydrodynamics in Early Time Relaxation}

\begin{figure}[h!]
\centering
\includegraphics[width=\textwidth]{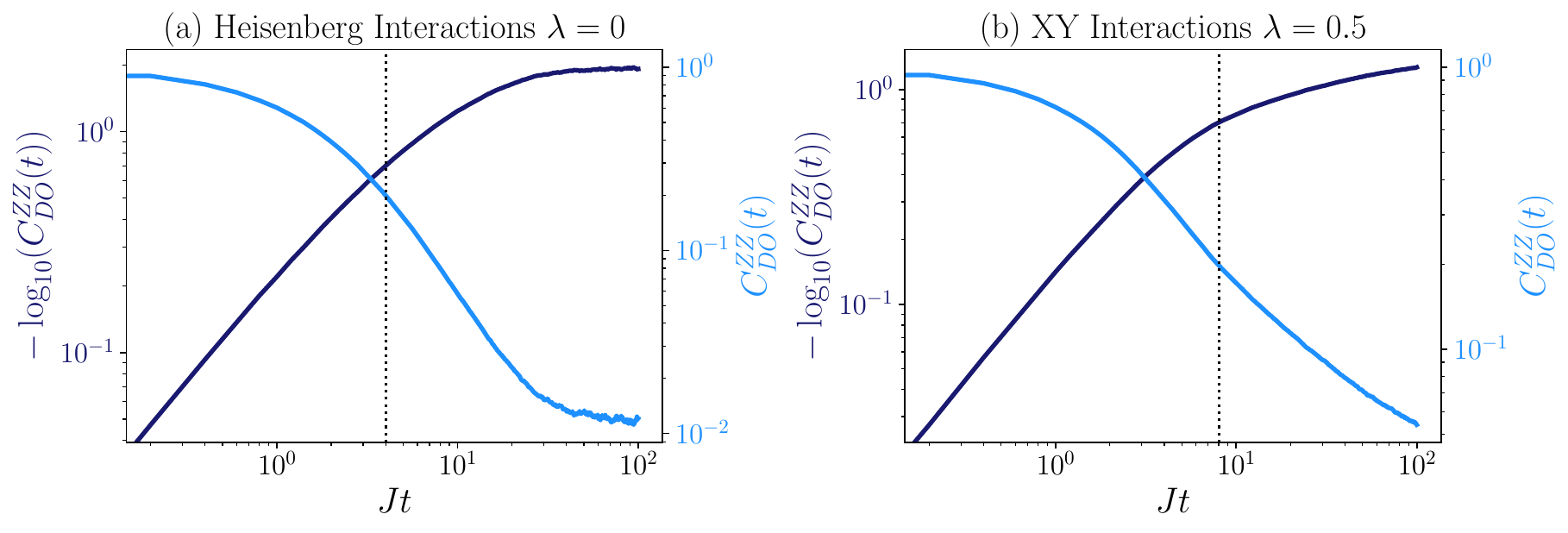}
\caption{\textbf{Coexistence of stretched exponential decay and hydrodynamics}: DTWA simulations out to long times of local Z autocorrelations, which theoretically should appear algebraic at late times when classical hydrodynamics is valid. Autocorrelations  for  Heisenberg (a) and XY (b) are shown simultaneously on a triple logarithmic plot (left axis, dark blue) and an ordinary logarithmic plot (right axis, light blue), designed to distinguish stretched exponential and algebraic decay respectively. While the stretched exponential form is best at early times in the fitting window (left of the vertical dashed line), algebraic decay is more visible at later timescales.}
\label{fig:hydro}
\end{figure}

Since we are probing the local $Z$ autocorrelation across the family of $XXZ$ Hamiltonians, one might expect to see algebraic decay predicted by classical emergent hydrodynamics, in contrast to the stretched exponent ansatz studied in this work.
Similarly for $X$ autocorrelations for the Heisenberg Hamiltonian. Numerical simulations in dTWA suggest that these pictures are mutually compatible. In particular, as is shown in Fig. \ref{fig:hydro}, algebraic decay is only perceptible outside of the fitting window relevant to the decoherence timescales of our experiment (to the left of the dashed vertical lines in Fig. \ref{fig:hydro} ).

It is nonetheless an interesting question for future work to study how features of the emergent hydroydnamics can be tuned by the engineering of microscopic interactions demonstrated in this work, on systems with longer experimentally accessible timescales. 

\section{Minimal Models of Local Thermalization Dynamics}

\subsection{Overview and Modified XXZ Parameterization}

%
We proceed to develop a series of models that systematically improve their agreement with the experiment, and crucially allow us to isolate individual physical effects and their influence on the disordered thermalization dynamics. 
First, we consider a toy model of the bond-disordered quantum many-body system in which the dipolar interactions are replaced by i.i.d. zero-mean, normally-distributed random variables. 
This model can exactly be mapped onto the dynamics of a single spin impurity being driven by a fluctuating magnetic field, namely a dynamical mean field theory.
Remarkably this all-to-all coupled version of the XXZ quench reproduces the qualitative dip in stretching exponent observed in the experiment. 
Next, we phenomenologically generalize this model to incorporate the ambient dimensionality and long-range nature of the dipolar interactions into a self-consistent local dynamical mean-field framework to reincorporate locality into the description. 
Neither of these dynamical mean-field models accurately capture the hierarchy of $X/Z$ relaxation timescales observed in the experiment and the numerics documented in Sec. \ref{sec:local-benchmark}, however.
To remedy this last discrepancy we develop a novel cluster generalization of the previous dynamical mean-field model to incorporate coherent interactions non-perturbatively within clusters of spins. 


In contrast to the parameterization introduced in the main text $\bm g(\lambda_{XXZ})$, which was motivated by the experimental Floquet engineering protocol, we introduce an alternative parameterization of XXZ Hamiltonians in this section motivated by the physical regimes of local thermalization.
\begin{align}\label{eq:periodic-XXZ}
\bm{g}(\theta) = \left(\cos{\left(\theta-\pi/4\right)}, \cos{\left(\theta-\pi/4\right)}, -\sin{\left(\theta-\pi/4\right)} \right).
\end{align}
This parameterization has a number of important properties, which help highlight different thermalization mechanisms of interest:

\begin{itemize}
\item The easy-axis parameter space $\theta \in \left(-\pi/2 , 0\right) $ is symmetric to the easy-plane parameter space $\theta \in \left(0, \pi/2\right) $
\item The special Hamiltonians, Ising, Heisenberg, XY and dipolar, are equally spaced at respective points $\theta=-\pi/4, 0, \pi/4, \pi/2$
\item The infinite temperature correlation functions are $\pi$-periodic in $\theta$. In particular, as one applies the transformation $\theta \to \theta + \pi$,  the XXZ anisotropy is negated $\bm{g}(\theta) \to -\bm{g}(\theta)$ which effectively applies a time-reversal operation on correlation function. However, the Hamiltonian itself, Eq. \ref{eq:XXZ-Dipoles}, is symmetric under time-reversal so the auto-correlation function will be invariant.
\end{itemize}

\begin{figure}[h!]
\centering
\includegraphics[width=\textwidth]{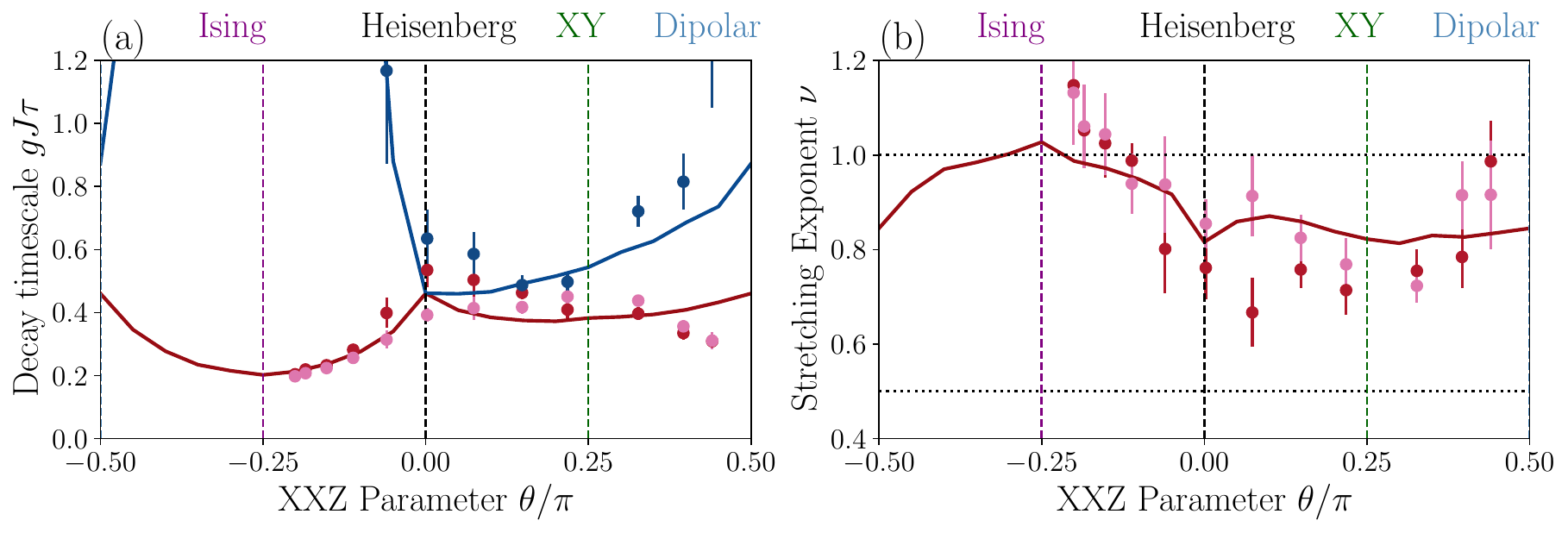}
\caption{\textbf{Local decay parameters in periodic parameterization}: Stretched exponential fit parameters (timescales (a) and stretching exponents (b)) extracted from $N=18$ Krylov simulations of the quench disordered XXZ model (thick lines), according to the periodic XXZ parameterization, Eq. \ref{eq:periodic-XXZ}. Experimental X (red circles), Y (pink circles) and Z (blue circles) data is recontextualized in this new parameterization. }
\label{fig:NewParam}
\end{figure}

The stretched exponential fit parameters extracted from the experiment together with those extracted from $N=18$ Krylov subspace calculations, are plotted in Fig.~\ref{fig:NewParam} in this new parameterization. Note that the experiment can only access the regime $\theta \in \left( -\pi/4, \pi/2 \right)$ via the Floquet engineering method and plotting in this periodic parameter space recontextualizes the previous dip feature of the $X$ stretching exponent as an \textit{oscillation} in the periodic XXZ phase diagram.

\subsection{Anistropic Sachdev-Ye Toy Model}


As an initial approach to understand the dynamical features observed in the experiment, we consider a toy XXZ model in which we discard the complex correlations originating from configurational disorder, and study the average dynamics when all couplings are taken as zero-mean, independent normally distributed random variables each with common variance,
\begin{align}\label{eq:SY}
H^{\text{Toy}}_{\bm{g}} = \sum_{ij} J_{ij} \sum_\mu g_\mu S_i^\mu S_j^\mu, \qquad \overline{J_{ij}}=0, \qquad \overline{J_{ij}^2} = \frac{2J^2}{N}. 
\end{align}
This is an anisotropic generalization of the Sachdev-Ye model, which can be mapped exactly onto a \textit{local} dynamical mean-field description in the limit of $N \to \infty$ \cite{sachdev1993gapless}. 
Rather than construct an approximate impurity solver~\cite{zhou2021disconnecting} to solve this dynamical mean-field theory, we resort to extracting the mean-field dynamics indirectly through $N=18$ Krylov simulations of the toy model itself, which is a more accurate description of the classical saddle point. 

\begin{figure}[h!]
\centering
\includegraphics[width=\textwidth]{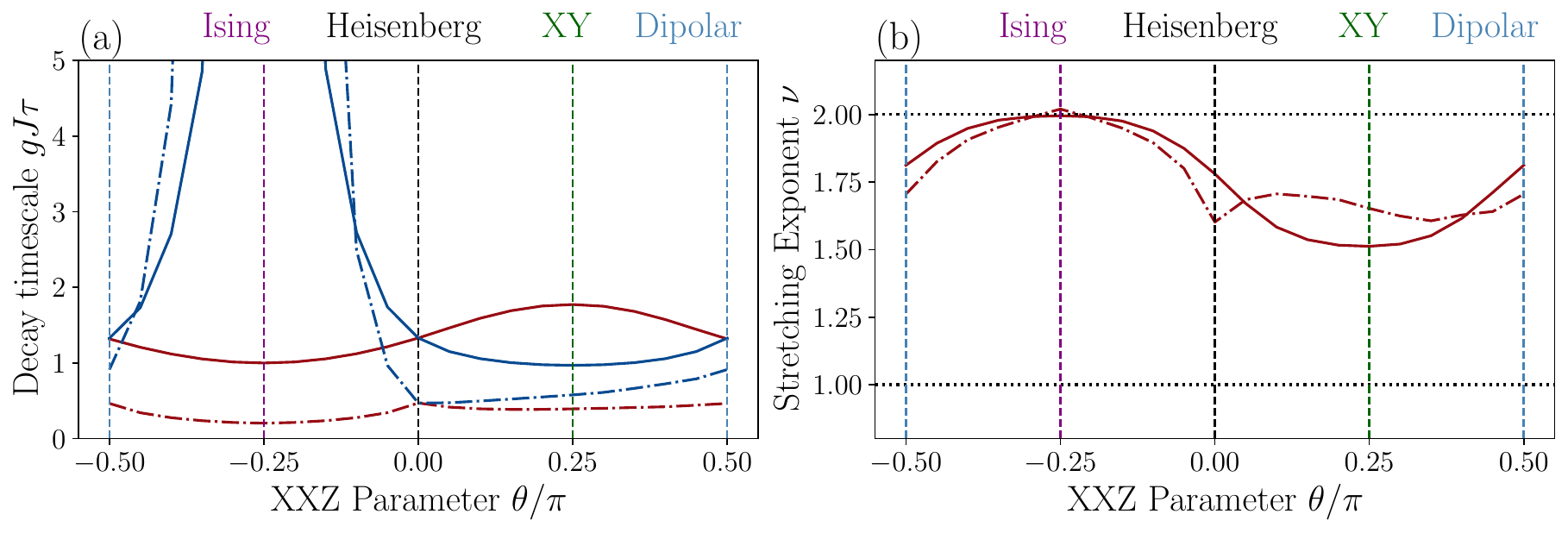}
\caption{\textbf{Stretched exponential relaxation of toy SY model}: Stretched exponential fit parameters (timescales (a) and stretching exponents (b)) extracted from $N=18$ Krylov simulations of the SY model (thick lines), Eq.~(\ref{eq:SY}). Analogous parameters extracted from the $d=\alpha=3$ dipolar calculations are plotted in dot-dashed lines, with stretching exponents rescaled by $2\alpha/d=2$ to agree with toy model.}
\label{fig:RSM}
\end{figure}

%
Using our stretched exponential ansatz again with a fitting threshold of $C_{\text{min}}=0.2$, we plot the extracted fit parameters in Fig.~\ref{fig:RSM}. 
Remarkably, the variation of the stretching exponent across the XXZ phase diagram observed in the experiment is reproduced, but now with shifted values, $\nu_{SY} \leq 2$, which is saturated for the Ising interaction. 
This limiting value can be understood analytically,
\begin{align}
C^{XX}_{Ising}(t) &= \overline{\prod_j \cos{J_{ij} t}} \\
&= \left(\exp{-\frac{1}{2}\text{Var}(J_{ij})t^2}\right)^{(N-1)} \\
&= \exp{-J^2 t^2 (1-1/N)} \to \exp{-(Jt)^2}.
\end{align}
We can physically understand this result in the same way as the Ising Hamiltonian in the main text: due to the local magnetization conservation law, the field acting on the classical impurity is static, inducing ballistic phase accumulation.
Note however the distinction from the $d/\alpha$ maximal stretching exponent obtained from realistic disorder averaging: the SY model does not have any notion of locality, and can be regarded as all-to-all connected limit of the realistic configurationally disordered result. 
Hence as $N \to \infty$, we are left with a single collective spin impurity driven by a collective bath, resulting in rapid Gaussian decay when the bath drives ballistic dynamics. 
Furthermore, when the collective bath acquires a dynamical character away from this integrable point, the finite correlation time of the magnetic field transverse to the $X$ axis reduces the stretching exponent of $X$ decay. This is observed in Fig. \ref{fig:RSM}(b) as one crosses into the easy-plane regime, leading to a dip of the stretching exponent at the $XY$ Hamiltonian, reflecting the sub-ballistic dephasing dynamics sourced by the $Y$ field's fluctuations.

%
This simplistic toy model still differs from our experimental observations in two important aspects:
\begin{itemize}
\item Experimental stretching exponents $\nu_{Exp}$ do not exceed $1= d/\alpha$.
\item $\tau_Z \geq \tau_X$ in experiment, in contrast with the crossover in timescales observed on the easy-plane side of the SY model.
\end{itemize}
The first point is remedied with a phenomenological mean-field model described in the following section, Sec. \ref{sec:DMFT}, which properly takes into account dipolar configurational averaging. 
The latter is addressed in Sec. \ref{sec:cDMFT} through the introduction of coherent interactions within clusters of spins.

\subsection{Dynamical Mean-Field Theory}\label{sec:DMFT}


Motivated by the success of the previous toy model in capturing the variation of the $X$ stretching exponent across the XXZ phase diagram, we make phenomenological modifications to incorporate the ambient dimensionality and spatial range of couplings. 
In particular, we consider a minimal local dynamical mean-field framework, in which the decay of each spin is driven by a fluctuating magnetic field $B_i^\mu$. 
We approximate the quantum fluctuations of magnetic field operators
\begin{align}
B^\mu_i = g^\mu \sum_j J_{ij} S^\mu_j,
\end{align}
by stochastic, normally-distributed classical fields $\lbrace \bm b_i(t) \rbrace_i $, characterized by the moments
\begin{align}
\overline{b_i^\mu(t)} = 0, \qquad \overline{b_i^\mu(t) b^\nu_j(t')} = \delta^{\mu \nu} \delta_{ij} D_i^\mu(t-t'),
\end{align}
which are consistent with infinite temperature quantum expectation values of the XXZ model~\cite{graer2021dynamic}. 
The noise kernel $D^\mu_i(t)$, describing the local field fluctuations which dephase the $i$th spin, are then linked to the dynamics of the rest of the spins in the ensemble through the self-consistency condition 
\begin{align}
D_i^\mu(t) = \left(g^{\mu}\right)^2 \sum_j J_{ij}^2 C_j^\mu(t),
\end{align}
where $C_j^\mu(t) = \overline{S^\mu_j(t)S^\mu_j(0)}$, is the spin-autocorrelation induced by the local dephasing generated by $D_j$. 
Given each geometry and coupling matrix $\lbrace J_{ij} \rbrace$, this self-conisistent problem is solved iteratively, sampling over $1000$ instances of local noise and iterating until a local tolerance of $10^{-2}$ is attained, quantified by the $L^2$ distance on the autocorrelation time-series
\begin{align}
d(C_1, C_2) = \frac{1}{T} \int_0^T dt' |C_1(t')-C_2(t')|^2.
\end{align}
Final results stretching exponents attained from this simulation, averaged again over $M=100$ geometric disorder realizations, are shown in Fig.~\ref{fig:DMFT-cDMFT}(d) as dashed lines as well as in Fig. 3 (b) of the main-text. 
Clearly this modified model successfully rescaled the stretching exponent values to lie between $\lbrack 1/2,1 \rbrack $, as opposed to the prediction of our previous toy model.
%


The intuition why this phenomenological modification of the previous toy model has rescaled the stretching exponent from between $\lbrack 1,2 \rbrack $ to $\lbrack d/2\alpha ,d/\alpha \rbrack $ is simple.
In the previous toy model, there was no notion of dimensionality or ambient connectivity of the spin bath; 
Spins were all to all coupled, giving a \textit{bare} stretching exponent determined purely relative decay to fluctuation timescales of the local impurity relaxation to lie between $\lbrack 1,2 \rbrack$. 
However, now in the phenomenological model the number of spins that effectively contribute to the dephasing of a given neighbour is constrained by the ambient dimensionality and coupling range, quantified by the effective \textit{bath connectivity} $d/\alpha$. 
Indeed, this geometric factor comes very simply from counting the number of spins that actually contribute to the dephasing dynamics as is explained in the main-text.


While the results (dashed lines in Fig.~\ref{fig:DMFT-cDMFT} (c,d)) from this simplistic minimal model correctly describe the qualitative behaviour of the $X$ stretching exponent, they incorrectly predict that the $Z$ degree of freedom decays faster than $X$ on the easy-plane side of the XXZ phase diagram (timescale ratio peak in Fig.~\ref{fig:DMFT-cDMFT} (c)), as the effective transverse field is larger for the $Z$ degree of freedom. 
In the subsequent section, we show that the observed absence of this time-scale cross-over is indicative of coherent hybridization among strongly coupled spins contributing to the dynamics in a way which cannot be captured by a naive dynamical mean-field theory.

\subsection{Role of Coherent Interactions}\label{sec:cDMFT}

In both the SY model and the  phenomenological dynamical mean-field model, the ratio of $X$ to $Z$ decay timescales $r=\tau_X/\tau_Z$ is severely overestimated. 
We proceed to provide analytical evidence that $r \leq 1$ is a signature of coherent hybridization between clusters of spins, and adapt our dynamical mean field theory to incorporate and verify these effects.

\subsubsection{Motivation: Pair Spin Dynamics}
Focusing for a moment on the case of two interacting spins, we proceed to demonstrate how the exact hybridization of the pair, together with the heavy-tailed probability distribution of their interaction energy imparted by positional disorder, conspire to yield commensurate X and Z timescales for all easy-plane Hamiltonians, $r_{\text{2 spins}}=1$. 
\begin{figure}[h!]
\centering
\includegraphics[width=\textwidth]{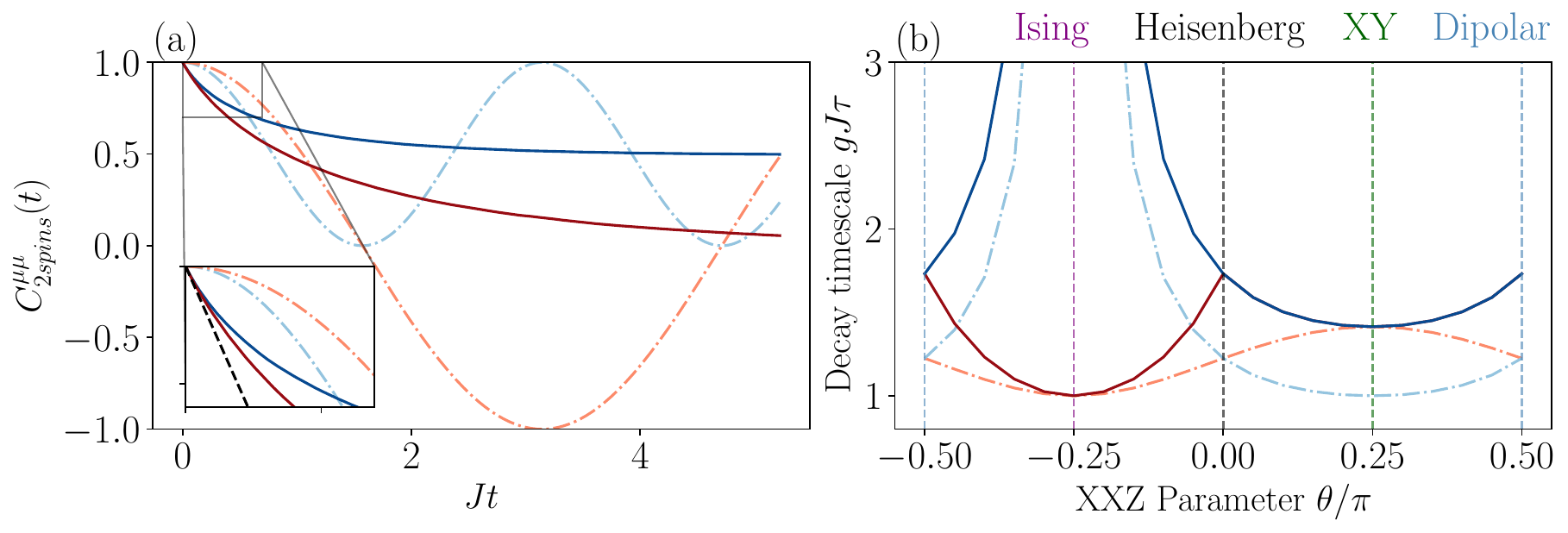}
\caption{\textbf{Pair spin dynamics}: (a) XY dynamics of two spins at a typical separation (dashed lines) compared against the disorder averaged autocorrelation (solid lines). Inset zooms into the early time regime, featuring the analytical early time result (black dashed line) in Eq.~(\ref{eq:disavg-earlytime}). Blue traces represent local $Z$ autocorrelations while red traces represent local $X$ autocorrelations. (b) Comparison of early time fit of respective curves from (a), across the XXZ phase diagram. By incorporating coherent effects captured by the disorder averaged timescale, the X/Z decay timescales of the pair become commensurate on the easy-plane side of the phase diagram.}
\label{fig:pairspins}
\end{figure}
First, let us calculate the exact autocorrelation for a generic $XYZ$ interaction between a pair of spins
\begin{align}\label{eq: exact_pair_XXZ}
C^{\mu\mu}_{\text{2 spins}}(t) &= \overline{\frac{1}{2}\sum_i \text{Tr}\left(e^{i H_{\bm g} t} \sigma_i^\mu e^{-i H_{\bm g} t}\sigma_i^\mu\right)} \\
&= \overline{\prod_{\nu \neq \mu} \cos{\left(g_\nu J(\bm r) t\right)}},
\end{align}
where the disorder average is taken over independent normal distributions of spin positions, with details given in Sec. \ref{sec: pairspin_calc}.
Taking the naive early-time expansion of this quantity (see Sec. \ref{sec:global}) yields the timescale 
\begin{align}
\tau_0^\mu  &= 1/\sqrt{\overline{J^2(\bm r)}\left( \sum_{\nu \neq \mu} g_\nu^2 \right)},
\end{align}
which is physically a measure of the field strength transverse to the $\mu$th direction, rescaled by the coupling strength to the spin-bath.  
The naive conclusion would then be that the timescale ratio is
\begin{align}\label{eq:naive-ratio}
r_0 &=\frac{\tau_0^X}{\tau_0^Z} = \sqrt{\frac{2}{1+ g_z^2/g_x^2}},
\end{align}
which is plotted as a function of the XXZ parameter in Fig.~\ref{fig:pairspins}(c) as a dot-dashed line, clearly predicting a maximum ratio of $\sqrt{2}$ at the XY point in agreement with our previous mean field models based upon precisely this second cumulant of the mean-field.

What this calculation misses, however, is the effect of the subsequent dynamics of both $X$ and $Z$ autocorrelations, which influence the decay strongly interacting spins even at early times, when we perform the average over positional disorder. 
In particular, the $Z$ autocorrelator oscillates about $1/2$ for all XXZ Hamiltonians (blue dashed lines in Fig.~\ref{fig:pairspins}), as is necessary due to the global $Z$ conservation law for this pair of spins. 
On the other hand, the $X$ degree of freedom oscillates around $0$ at late time, as it is not conserved (red dashed lines in Fig.~\ref{fig:pairspins}). 
Thus, despite the fast early time decay of $Z$ in the easy plane regime compared to $X$, its conservation at late time should \textit{elongate} its relaxation time in the disorder averaged trace (solid lines in Fig.~\ref{fig:pairspins}(a)).
In fact, analytically incorporating the positional disorder \textit{before} taking the early-time limit correctly predicts that $X/Z$ timescales are commensurate in the easy-plane regime (see Fig. \ref{fig:pairspins}(b) solid lines) 
\begin{align}
C_{\text{2 spins}}^{\mu\mu}(t) = 1-|t|/\tau_1^\mu +\mathcal{O}((Jt)^2), \qquad t \to 0,
\end{align}
where
\begin{align}\label{eq:disavg-earlytime}
\left(J \tau_1^\mu\right)^{-1} = \frac{32}{9 \sqrt{3}}\left(|g^\mu_{\perp,1} + g^\mu_{\perp,2}| + |g^\mu_{\perp,1} - g^\mu_{\perp,2}|\right)
\end{align}
is written in terms of the two couplings transverse to $\mu$, $g^\mu_{\perp,i}$. See Sec. \ref{sec: pairspin_calc} for a detailed derivation of this result. 

Crucially, this quantity is maximized at fixed Hamiltonian norm only if $g^\mu_{\perp,i} = \pm g^\mu_{\perp,i}$, which is the same condition as there being a conservation law for the pair of spins ($S_1^\mu \pm S_2^\mu$ is a conserved charge in each case). 
This effect is visible in Fig.~\ref{fig:pairspins}(b), where we see the maximum deviation from the naive X timescale prediction Eq.~(\ref{eq:naive-ratio}) at the Heisenberg ($XX+YY+ZZ$) and anti-Heisenberg ($XX+YY-ZZ$) Hamiltonians. 
\subsubsection{Cluster Dynamical Mean Field Theory (cDMFT)}

Having understood the importance of coherent interactions and conservation laws in determining the ensemble averaged timescales, we proceed to introduce a cluster-based generalization of the dynamical mean field framework constructed in Sec.~\ref{sec:DMFT}.
In particular, we will start by dividing the geometrical spin configuration into \textit{clusters} of strongly coupled spins via a natural clustering algorithm. Given a clustering threshold $J_0$, any pair of spins with coupling strength exceeding the threshold $(|J_{ij}| \geq J_0)$ will be grouped into the same cluster.
When this cluster partition is determined, the Hamiltonian can be rewritten exactly as 
\begin{align}
H &= \sum_{a} H_a, \\
H_a &= \sum_{ij \in a}J_{ij} g_\mu S_i^\mu S_j^\mu + \sum_{i \in a}S_i^\mu \sum_b \sum_{j \in b} J_{ij} g_\mu  S_j^\mu,
\end{align}
where $a,b$ index the clusters. 
To gain a better approximation of the interacting dynamics, we proceed to approximate all inter-cluster interactions with classical fluctuating fields, while treating intra-cluster interactions coherently
\begin{align}\label{eq:cluster-ham}
H_a &\approx \sum_{ij \in a} g_\mu S_i^\mu S_j^\mu + \sum_{i \in a} b^\mu_i(t) S_i^\mu \qquad \overline{b_i^\mu(t) b_j^\nu(t')} \approx \delta_{ij} \delta^{\mu \nu} \left(D^a\right)_i^\mu(t-t'),
\end{align}
where we invoke the self-consistency condition
\begin{align}
\left(D^a\right)_i^\mu(t) &= \left(g^{\mu}\right)^2 \sum_b \sum_{j \in b} J^2_{ij}  \left(C^b\right)^{\mu}_{j}(t),
\end{align}
with
\begin{align}
\left(C^b\right)^\mu_{j}(t) &= \overline{\text{Tr}\left( S_j^\mu(t) S_j^\mu(0) \right)/ \text{Tr}\left(\bm{1}_b \right) }
\end{align}
calculated exactly with Krylov subspace techniques for trotterized time-evolution under the stochastic cluster Hamiltonians, Eq. \ref{eq:cluster-ham}.
It is straightforward to notice that in the limit where the clusters are simply single spins, this model is exactly equivalent to the naive dynamical mean-field approach described in the previous section.

\begin{figure}[h!]c
\centering
\includegraphics[width=\textwidth]{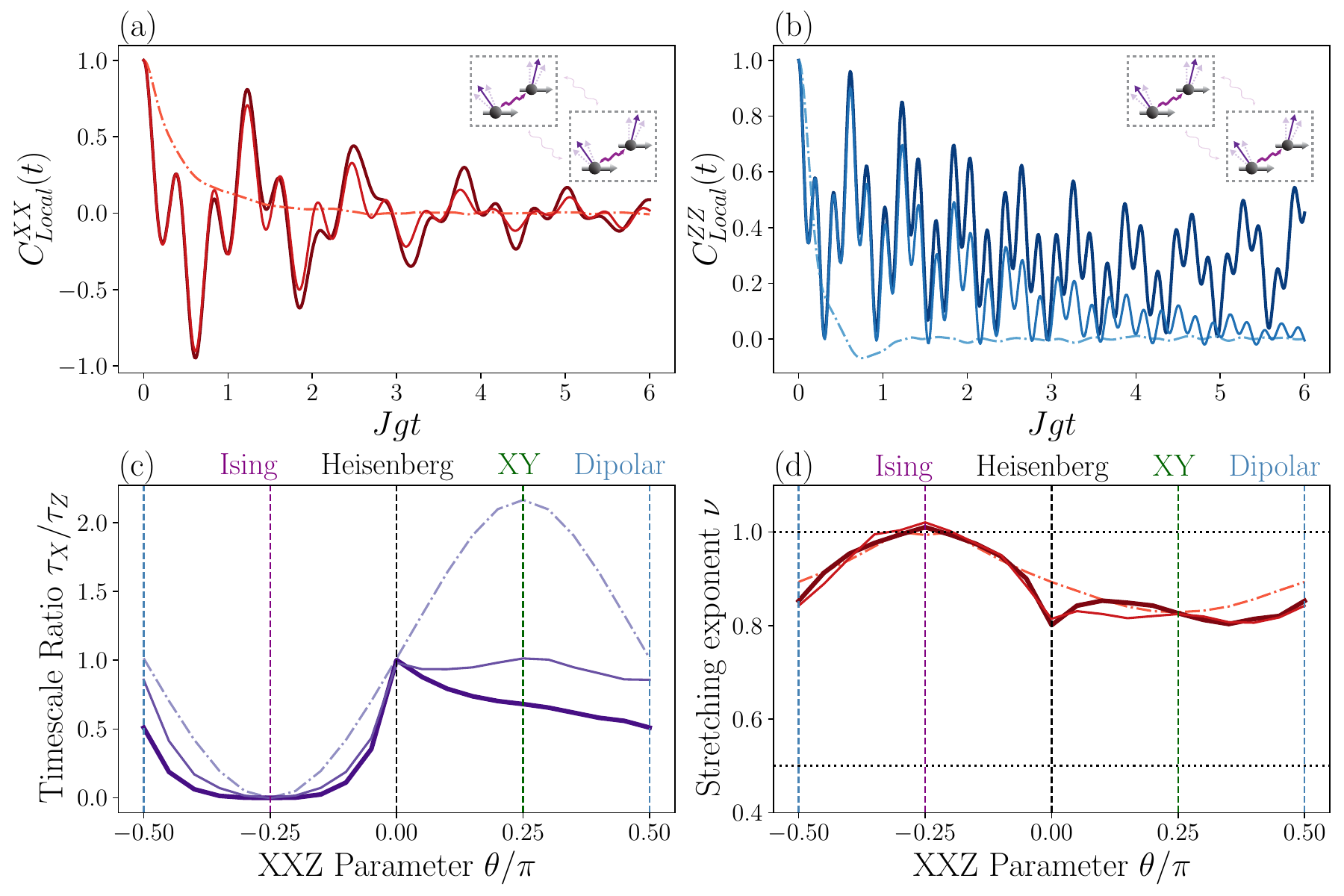}
\caption{\textbf{cDMFT and DMFT Predictions}: Comparison of exact XY dynamics (dark thick line) for X (a) and Z (b) local autocorrelators of the quadruplet of spins in the fixed disorder instance (inset) against predictions from the DMFT (dashed thin line) and cDMFT (solid thin line) predictions.  cDMFT clearly extends the lifetime of conserved quantities via its inclusion of coherent back-action within clusters.  (c,d) Disorder averaged comparison at fixed system size $N=18$, among exact Krylov (thick solid line), cDMFT (thin solid line) and DMFT (thin dashed line) for predicted timescale ratios and stretching exponent.}
\label{fig:DMFT-cDMFT}
\end{figure}

Before comparing this algorithm to Krylov calculations for quantities measured in the experiment, we first recapitulate the intuition presented in the previous section regarding conservation laws for a quadruplet of spins which are grouped into two weakly coupled pairs of spins, as depicted in the insets of Fig.~\ref{fig:DMFT-cDMFT} (a,b).
For the exact local $X$ correlator, the system size is still small enough to see a full oscillation from $1$ to $-1$ in Fig.~\ref{fig:DMFT-cDMFT} (a) due to its unrestricted motion from a lack of global conservation.
The $Z$ degree of freedom's autocorrelations, just as in the two spin example, are constrained to oscillate between $1$ and $0$ due to its global conservation. 
While the dynamics of the generic local observable $X$ is well-described by both methods, the dynamics of the conserved observable $Z$ improves substantially with the cluster approach, as is observed in Fig. \ref{fig:DMFT-cDMFT}(a,b).

Finally we compare both DMFT and cDMFT to our Krylov simulations results in Fig.~\ref{fig:DMFT-cDMFT} (c,d) for a clustering threshold value of $J_0 = 1.75 J$. 
The timescale ratio is significantly reduced on the easy-plane side of the phase diagram, again confirming this signature of conservation laws as claimed in the main-text. 
The conservation laws additionally have an impact on the stretching exponents of producing a strong dip feature at the Heisenberg Hamiltonian, precisely due to the SU(2) symmetry and associated $X$ conservation law, stretching the decay shape as expected. 
The fact that this dip is missed by the dynamical mean-field model is additional evidence of this interpretation.  

In summary, we have established analytical and numerical evidence that coherent interactions and conservation laws provide concerete signatures in the ideal XXZ local correlation functions that are witnessed in the experimental data.

\subsubsection{Analytical Calculations for Pair Spin Dynamics}\label{sec: pairspin_calc}

From Eq. \ref{eq: exact_pair_XXZ}, the exact two-spin autocorrelations of the XXZ model at infinite temperature are 

\begin{align}
C^{XX}_{\text{2 spins}}(t) &= \overline{\cos{\left(g_x J t\right)}\cos{\left(g_z J t\right)}}, \\
C^{ZZ}_{\text{2 spins}}(t) &= \overline{\cos{\left(g_x J t\right)}^2}.
\end{align}

It is most convenient to analyze the early time behaviour of this quantity via the form
\begin{align}
C^{XX}_{\text{2 spins}}(t) &= 1-F((g_x-g_z)t/2)-F((g_x+g_z)t/2),\\
C^{ZZ}_{\text{2 spins}}(t) &= 1-F(g_xt),
\end{align}
where
\begin{align}
F(\tau) &=  \overline{\sin^2\left(J \tau \right)},
\end{align}
intuitively represents the averaged coherent exchange between the pair. For analytical convience we assume each spin position is independently drawn a gaussian probability distribution of zero mean and variance $L$, a length-scale that sets a typical energy scale that we will calculate shortly. We proceed to  calculate 

\begin{align}
F(\tau) &= \int \frac{d\bm r_1}{\left(2\pi \right)^{d/2}L^d}\frac{d\bm r_2}{\left(2\pi \right)^{d/2}L^d} \, \sin^2\left(J(\bm r_{12}) \tau \right)\exp{-\frac{1}{2L^2}\left(r_1^2+r_2^2\right)} \\
&= \frac{1}{\left(2\pi \right)^dL^{2d}} \left(\int d\bm R_{12} e^{-R_{12}^2/L^2}\right) \int d\bm r_{12} \exp{-\frac{r_{12}^2}{4L^2}}\, \sin^2\left(J(\bm r_{12})\tau \right) \\
&= \frac{\pi^{d/2}L^d}{\left(2\pi \right)^dL^{2d}} \, \int d\Omega \int_0^\infty \frac{dr}{r} r^d\, \exp{-\frac{r^2}{4L^2}} \, \sin^2\left(q(\Omega) \tau \, r^{-\alpha}\right) \\
&= |\tau|^{d/\alpha}\frac{\pi^{d/2}L^d}{\left(2\pi \right)^dL^{2d}} \frac{1}{\alpha} \, \int d\Omega |q(\Omega)|^{d/\alpha}  \int_0^\infty \frac{dz}{z^{1+d/\alpha}} \, \sin^2{z} \, \exp{-\frac{\left(|q(\Omega)\tau| /z\right)^{2/\alpha}}{4L^2}} \\
&= \frac{1}{\alpha}|\tau|^{d/\alpha} \frac{1}{2^d\pi^{d/2}L^d}|| q ||_{d/\alpha}^{d/\alpha} \, \int_0^\infty \frac{dz}{z^{1+d/\alpha}} \sin^2z \exp{-\frac{\left(|q(\Omega)\tau| /z\right)^{2/\alpha}}{4L^2}} \\
&\approx \frac{1}{9}\sqrt{\frac{\pi}{3}} \frac{|\tau|}{L^3} \qquad \tau \to 0, d=\alpha=3
\end{align}

where we introduced $\bm r_{12}=\bm r_1-\bm r_2, \bm R_{12}=(\bm r_1+\bm r_2)/2$, calculated the $L^1$ norm of the angular coupling

\begin{align}
||q||_1 = 2\pi \int_{-1}^{1} dx |3x^2-1| = \frac{16 \pi}{3 \sqrt{3}},
\end{align}

and also considered the change of variables $z= q(\Omega) \tau r^{-\alpha}$, $dz/z = \alpha dr/r$. Lets also consider how to define a typical energy scale of the pair. The typical interspin distance is 

\begin{align}
a &=\overline{r_{12}} \\
& = \int \frac{d^dr_1}{\left(2\pi \right)^{d/2}L^d}\frac{d^dr_2}{\left(2\pi \right)^{d/2}L^d} \, r_{12} \, \exp{-\frac{1}{2L^2}\left(r_1^2+r_2^2\right)} \\
&= \frac{1}{\left(2\pi \right)^dL^{2d}} \left(\int d^dR_{12} e^{-R_{12}^2/L^2}\right) \int d^dr_{12} \exp{-\frac{r_{12}^2}{4L^2}}\, r_{12} \\
&= \frac{\pi^{d/2}L^d}{\left(2\pi \right)^dL^{2d}} \, \int d\Omega \int_0^\infty dr \, r^{d}\, \exp{-\frac{r^2}{4L^2}} \\
&= \frac{\pi^{d/2}L^d}{(2\pi)^d L^{2d}} L^{d+1}2^d \Gamma\left(\frac{d+1}{2}\right) S_d \\
&= 4 L/\sqrt{\pi} \qquad d=\alpha =3
\end{align}

We then define the relevant exchange rate at this length scale $J = 1/a^3 = \pi \sqrt{\pi}/4^3 L^3$. This gives the final result 

\begin{align}
    F(\tau) &= \frac{64}{9 \sqrt{3}}J |\tau|,
\end{align}

at early time from which Eq.~\ref{eq:disavg-earlytime} directly follows.

\bibliography{main}